\documentclass[12pt]{article}
\usepackage{amsfonts}
\usepackage{bbm}
\usepackage{booktabs}
\usepackage{graphicx}
\usepackage{mathrsfs}
\usepackage{color}
\usepackage{verbatim}
\usepackage{cancel}
\usepackage{subfigure}
\usepackage{algpseudocode}
\usepackage{amsmath}
\usepackage{setspace}
\usepackage{geometry}             
\usepackage{amssymb}
\usepackage{amsmath}
\usepackage{epstopdf}
\usepackage{cases}
\usepackage{cite}

\usepackage[hyperindex=true,
          pdfstartview=FitH,
          bookmarksnumbered=true,
          bookmarksopen=true,
          citecolor=blue,
          linkcolor=blue,
          colorlinks=true,
          urlcolor=rossoCP3,
          unicode]{hyperref}

\definecolor{rossoCP3}{cmyk}{0,.88,.77,.40}

\begin{document}

\title{\bf Probabilistic Evolution of Black Hole Thermodynamic States via Fokker-Planck Equation}
 \author{\small Chao Wang${}^{1}${}, Chen Ma${}^{2}${} , Meng-Ci He${}^{1}${} and Bin Wu${}^{2,3,4,5}$\thanks{{\em email}: \href{mailto:binwu@nwu.edu.cn}{binwu@nwu.edu.cn}}{ }
	\vspace{5pt}\\
 	\small $^{1}${\it Institute of Physics, Shaanxi University of Technology, Hanzhong 723000, China}\\
 	\small $^{2}${\it School of Physics, Northwest University, Xi'an 710127, China}\\
 	\small $^{3}${\it Shaanxi Key Laboratory for Theoretical Physics Frontiers, Xi'an 710127, China}\\
 	\small $^{4}${\it Peng Huanwu Center for Fundamental Theory, Xi'an 710127, China}\\
 	\small $^{5}${\it Fundamental Discipline Research Center for Quantum Science and technology of }\\
 	\small {\it Shaanxi Province, Xi’an 710127, China}
 }

\date{}
\maketitle

\begin{spacing}{1.2}
\begin{abstract}
Employing the generalized free energy landscape and solving the associated Fokker-Planck equation, we obtain the time-dependent probability evolution of the order parameter for the RN-AdS black hole phase transitions. Our analysis reveals two distinct kinetic regimes, namely relaxation dynamics initialized at the unstable maximum and phase transition from the metastable state. Furthermore, we characterize the non-equilibrium irreversibility and macroscopic uncertainty using the entropy production rate and the Shannon entropy. The results demonstrate that the phase transition synchronizes exactly with a prominent peak in the entropy production rate, identifying the barrier crossing event as a process fundamentally driven by maximum thermodynamic dissipation.
\end{abstract}

\section{Introduction}

The historical development of thermodynamics reveals a universal trajectory, transitioning from equilibrium to non-equilibrium states and from deterministic to stochastic descriptions. This evolution finds a striking parallel in the study of gravitational systems, particularly black holes. The conceptual foundation of black hole thermodynamics was established by attributing physical entropy and temperature to these singular objects, effectively bridging the gap between general relativity and statistical mechanics. These foundational insights, codified into the four laws of black hole mechanics, profoundly influenced theoretical physics and redefined our understanding of spacetime.

In the subsequent decades, the theoretical framework for equilibrium thermodynamics has been rigorously refined. Wald defined black hole thermodynamic varibles as the Noether charge\cite{Wald:1993nt}, providing a geometric foundation applicable to general gravity theories \cite{Iyer:1994ys,Jacobson:1993vj,Clunan:2004ch,Kim:2013cor}. The interpretation of the cosmological constant as pressure \cite{Kastor:2009wy,Dolan:2010ha,Dolan:2011xt} leads to the extended phase space formalism, which reveals that AdS black holes exhibit Van der Waals-like phase structures \cite{Kubiznak:2012wp}. More recently, motivated by the necessity of restoring thermodynamic extensivity \cite{Visser:2021eqk,Cong:2019fqn}, the restricted phase space formalism was established \cite{Gao:2021xtt}. In this paradigm, the central charge is introduced as a new thermodynamic variable while the geometric volume is kept fixed, thereby resolving the Euler relation scaling issue. This paradigm enriches the first law and provides a novel perspective on holographic duality, particularly in higher curvature theories like Gauss-Bonnet gravity \cite{Wang:2022err} and scenarios with variable Newton's constant \cite{Wang:2021cmz}.

To reveal the microscopic mechanisms underlying these transitions, recent research has shifted focus to non-equilibrium stochastic dynamics using the free energy landscape topography \cite{Li:2020khm,Li:2020nsy}. In this framework, the black hole phase transition is no longer a sudden jump but a continuous stochastic process driven by thermal fluctuations. By treating the order parameter as a stochastic variable governed by the Langevin and Fokker-Planck equations \cite{Risken:1989}, researchers can quantitatively analyze the probabilistic evolution of the system. This approach has proven powerful in calculating the mean first passage time (MFPT) for barrier crossing \cite{Wei:2021bdr,Li:2022khi}, exploring the kinetic paths via the Onsager-Machlup path integral formalism \cite{Li:2024path}, and identifying critical slowing down phenomena near critical points \cite{Yang:2025def}. Furthermore, this framework has been successfully applied to black holes confined in a cavity \cite{Li:2024cavity}, revealing that the kinetics of the phase transition are fundamentally controlled by the height of the free energy barrier. Extending beyond traditional phase boundaries, recent thermodynamic analyses have also mapped complex phase structures in the supercritical regime, delineating continuous crossovers and Widom lines governed by the properties of the Gibbs free energy \cite{Xu:2025jrk}. While these extensive studies have successfully mapped various kinetic behaviors and phase structures, the overarching continuous temporal evolution of the system requires further exploration.

To elucidate the uncertainty of black hole evolution arising from thermal fluctuations and to investigate how this uncertainty depends on temporal changes, we systematically investigate the probabilistic evolution of RN-AdS black hole phase transitions by solving the Fokker-Planck equation on the free energy landscape. We categorize the stochastic dynamics into three distinct regimes: kinetic trapping, phase transition, and unstable relaxation. Furthermore, we utilize the Shannon entropy to quantify the macroscopic uncertainty, and the entropy production rate to characterize the non-equilibrium irreversibility. A pivotal finding is that the critical moment of the phase transition exactly coincides with a prominent peak in the entropy production rate. This reveals that the barrier crossing event is fundamentally a highly dissipative process driven by thermal fluctuations.

This paper is organized as follows. In Section \ref{II}, we briefly review the thermodynamics of the RN-AdS black hole and construct the generalized free energy landscape. Section \ref{III} solves the Fokker-Planck equation to obtain the time dependent probability distribution of the black hole evolution. Section \ref{IV} quantifies the non-equilibrium irreversibility and dynamic uncertainty of the evolution process by calculating the Shannon entropy and the entropy production rate. Finally, Section \ref{V} summarizes our conclusions.

\section{Set up}\label{II}

The free energy landscape serves as a fundamental framework for investigating phase transitions. By extending the thermodynamic description beyond equilibrium states, it provides a global perspective on stability and offers an intuitive topography for analyzing the stochastic evolution between different black hole phases.

\subsection{Reissner-Nordström-AdS Black Hole}

We consider the four-dimensional Reissner-Nordström anti-de Sitter (RN-AdS) black hole, governed by the Einstein-Maxwell-AdS action as
\begin{equation}
I =  \frac{1}{16 \pi G} \int d^4x \sqrt{-g} \left(R  + \frac{6}{l^2}  -F_{\mu \nu} F^{\mu \nu}  \right), \nonumber
\end{equation}
where $R$ is the Ricci scalar, $l$ denotes the AdS radius, and $G$ is Newton's gravitational constant. The electromagnetic field tensor is defined as $F_{\mu \nu} =\partial_\mu A_\nu - \partial_\nu A_\mu$, where the vector potential is $A_\mu=(-\Phi(r), 0,0,0)$. The static, spherically symmetric solution to the field equations is given by the metric below
\begin{equation}
\mathrm{d} s^2 = - f(r) \mathrm{d} t^2 + \frac{\mathrm{d} r^2}{f(r)} +r^2 \mathrm{d} \theta^2 + r^2 \sin^2 \theta \mathrm{d} \phi^2, \nonumber
\end{equation}
with the metric function $f(r)$ and the electrostatic potential $\Phi(r)$ taking the form of
\begin{equation}
f(r)=1-\frac{2 M}{r}+\frac{Q^2}{r^2}+\frac{r^2}{l^2}, \quad \Phi(r) = \frac{Q}{r}. \nonumber
\end{equation}
Here, $M$ and $Q$ represent the mass and charge of the black hole, respectively. The event horizon radius $r_h$ is determined by the largest root of $f(r_h)=0$. In terms of the horizon radius, the black hole mass (identified as the enthalpy $H$ in the extended phase space) is expressed as\cite{Kastor:2009wy}
\begin{equation}
M = \frac{r_h}{2}\left(1 + \frac{8}{3}\pi P r_h^2 + \frac{Q^2}{r_h^2}\right), \nonumber
\end{equation}
where the cosmological constant corresponds to the thermodynamic pressure $P = -\Lambda/(8\pi) = 3/(8\pi l^2)$\cite{Kubiznak:2016qmn}. The associated Hawking temperature $T_H$ and the Bekenstein-Hawking entropy $S$ are given by the standard relations\cite{Kubiznak:2012wp}
\begin{equation}
T_H = \frac{f'(r_h)}{4\pi} = \frac{1}{4\pi r_h}\left(1 + 8\pi P r_h^2 - \frac{Q^2}{r_h^2}\right), \quad S = \pi r_h^2. \nonumber
\end{equation}

\begin{figure}[!ht]
\centering
\includegraphics[width=8cm]{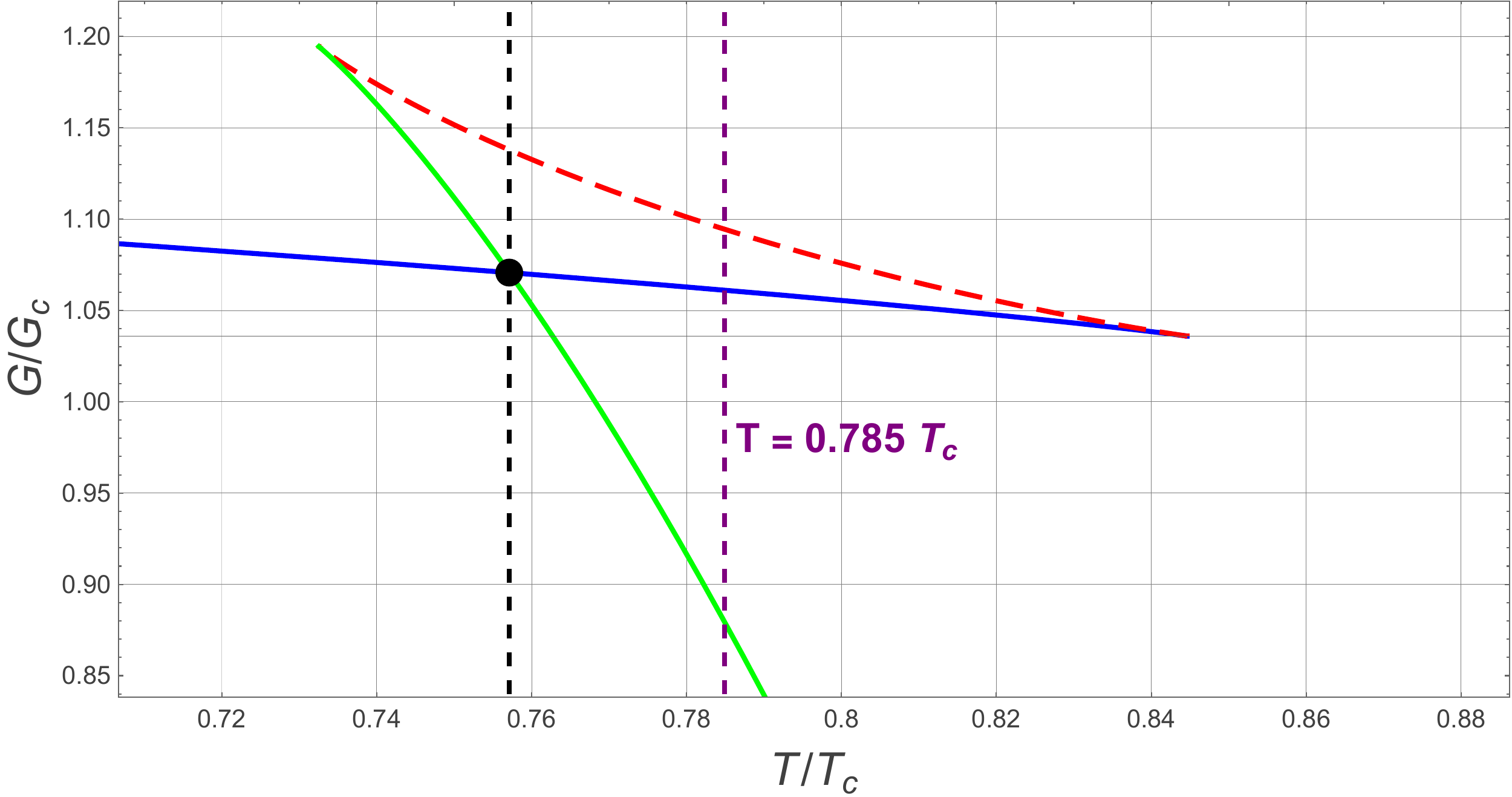}	
\caption{Isobaric Gibbs free energy $G$ versus temperature $T$ for pressure $P =0.5 P_c$. 
}\label{swallowtail}
\end{figure}

The RN-AdS black hole exhibits rich phase structures analogous to the liquid-gas phase transition in Van der Waals fluids. The critical point, determined by the inflection condition of the equation of state, i.e., $\left(\partial P / \partial r_h\right)_T=\left(\partial^2 P / \partial r_h^2\right)_T=0$, is explicitly given by 
\begin{equation}
	r_c=\sqrt{6} Q,\quad T_c=\frac{\sqrt{6}}{18 \pi Q},\quad  P_c=\frac{1}{96 \pi Q^2}.\nonumber
\end{equation} 
Without loss of generality, we set the black hole charge to unity ($Q=1$) in the following discussions. This behavior is visually captured by the characteristic swallowtail structure in the isobaric $G-T$ diagram with $G=M-T_HS$, as illustrated for a representative subcritical pressure $P=0.5 P_c$ in Fig.\ref{swallowtail}.
The blue solid, red dashed, and green solid lines represent the stable small black hole, the unstable intermediate state, and the stable large black hole branches, respectively. The system's equilibrium state strictly corresponds to the global minimum of the Gibbs free energy (i.e., the large black hole at the purple dashed line with $T=0.785T_c$). The intersection point of the SBH and LBH branches marks the coexistence state of the first order transition. Geometrically, the metastable and unstable branches terminate at specific cusps, indicating the exact thermodynamic points where the metastable state and the unstable intermediate state merge into a single inflection point. Macroscopically, as the temperature crosses this critical value, the black hole undergoes a dynamical transition in horizon radius to maintain the thermodynamically preferred state.

While the swallowtail diagram successfully identifies equilibrium states, it offers no insight into the dynamic pathway of the phase transition itself. To describe the stochastic evolution between phases, we employ the generalized free energy landscape\cite{Li:2020khm,Li:2020nsy}. This framework lies in treating the black hole as a thermodynamic system immersed in a canonical heat bath with a fixed temperature $T$ and pressure $P$. In thermal equilibrium, the black hole horizon radius is strictly denoted as $r_h$, where its intrinsic Hawking temperature matches the heat bath temperature ($T_H(r_h) = T$). To describe non-equilibrium processes, we introduce the horizon radius of an off-shell black hole as a dynamic order parameter $r$. The generalized free energy is then constructed as
\begin{equation}
\mathcal{G}(r; T, P) = M(r, P) - T \pi r^2 = \frac{r}{2}\left(1 + \frac{8}{3}\pi P r^2 + \frac{Q^2}{r^2}\right) - \pi T r^2.	\label{fre}
\end{equation}
The generalized free energy function $\mathcal{G}(r)$ delineates the energetic cost for the black hole to transition between different phases. To explicitly investigate the kinetics of this transition, we focus on the specific representative process at the aforementioned environment temperature $T=0.785 T_c$. 
\begin{figure}[!ht]
\centering
\includegraphics[width=8cm]{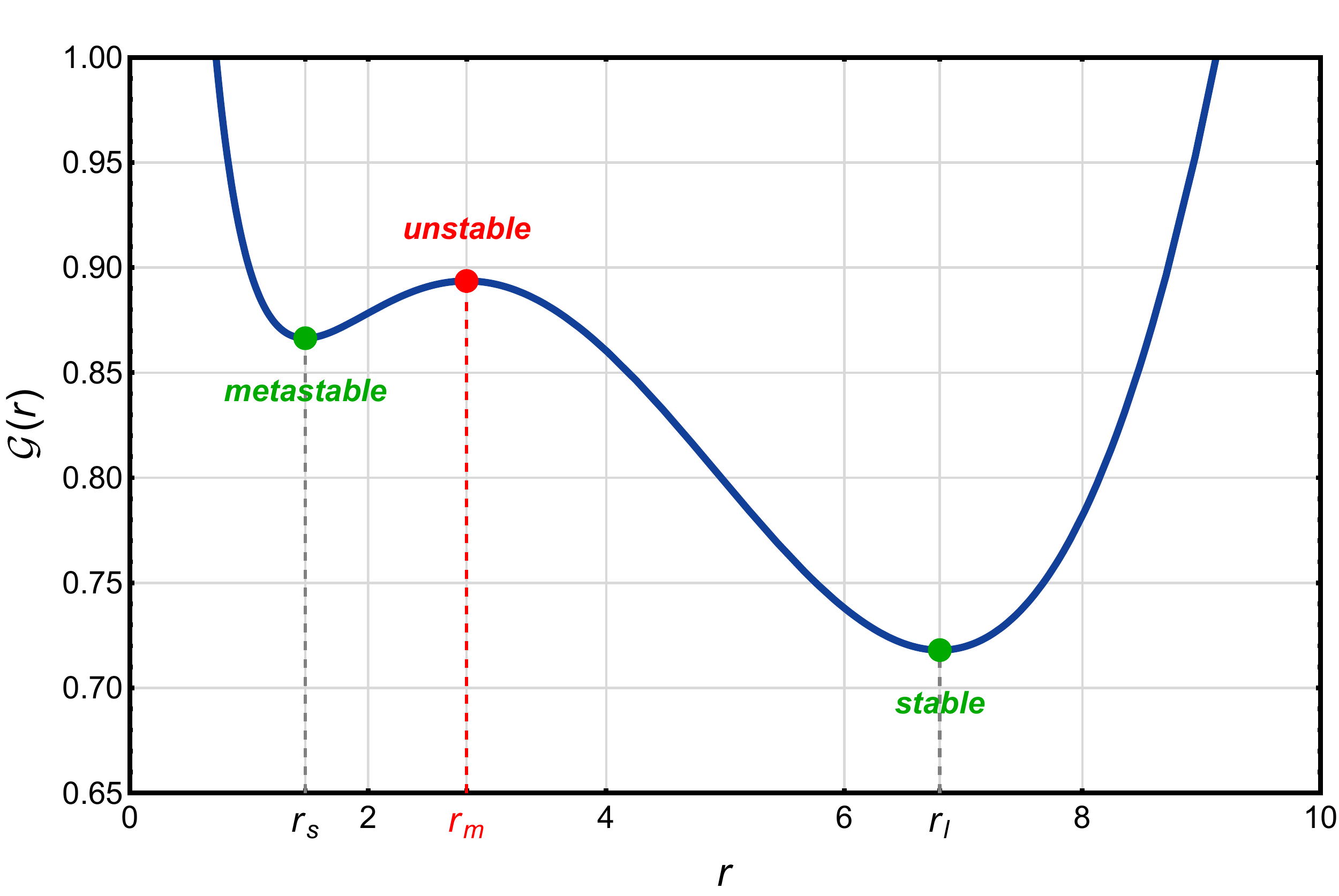}	
\caption{The generalized free energy landscape $\mathcal{G}(r)$ at $T=0.785 T_c$ (with fixed $P=0.5P_c$).}\label{landscape}
\end{figure}
As shown in Fig.\ref{landscape}, the generalized free energy landscape at this specific temperature well below the critical point ($T=0.785 T_c$) exhibits a characteristic double-well structure, where the two local minima correspond to the metastable SBH ($r_s$) and the stable LBH ($r_l$) states, separated by a local maximum representing the unstable intermediate black hole ($r_m$). This topographical framework allows us to dynamically trace how the black hole stochastically evolves toward lower free energy states driven by thermal fluctuations. Since our analysis focuses on stochastic transitions at a fixed temperature within the first-order regime, we present the landscape using the physical horizon radius $r$  rather than the reduced variable $r/r_c$
. This choice preserves the direct geometric meaning of $r$ and the explicit barrier cost. Nevertheless, if the goal were to study universal scaling laws near the critical point ($T \to T_c$), a dimensionless representation is mathematically straightforward.

\section{Stochastic Dynamics of Phase Transitions}\label{III}

To investigate the dynamic evolution of the black hole phase transition, we treat the horizon radius $r$ as a stochastic order parameter evolving under the influence of a thermal heat bath at temperature $T$. Analogous to a Brownian particle driven by microscopic collisions, the black hole undergoes continuous stochastic interactions with the reservoir. These thermodynamic perturbations induce fluctuations in the horizon radius, necessitating a probabilistic rather than deterministic description of the system's trajectory. To capture this stochasticity mathematically, we employ the Langevin equation for individual dynamic trajectories and the associated Fokker-Planck equation for the ensemble probability distribution.

In a full dynamical description, inertia would induce high frequency oscillations around the potential minima. However, our primary focus is on the long timescale probabilistic evolution towards equilibrium, rather than transient inertial relaxation. Since the characteristic timescale of the phase transition is significantly longer than the inertial relaxation time, the acceleration term becomes negligible. Consequently, the motion of the system in the generalized free energy landscape is governed by the overdamped Langevin equation \cite{ZNSM,Kampen},
\begin{equation}
	\frac{dr}{dt} = -\frac{1}{\zeta}\frac{\partial \mathcal{G}(r)}{\partial r} + \xi(t). \label{lang}
\end{equation}
Before detailing the specific terms, it is crucial to clarify the physical meaning of the time parameter $t$. In this context, $t$ is not the coordinate time of the static spacetime metric. Instead, it represents the macroscopic thermodynamic time of the external heat bath, parametrizing the kinetic relaxation process during which the black hole stochastically evolves toward equilibrium via continuous interactions with the reservoir. With this temporal framework established, in Eq.\eqref{lang}, the inertial term $\ddot{r}$ is omitted in the overdamped approximation, and $\zeta$ represents the friction coefficient characterizing the dissipation intensity. The first term $-\partial \mathcal{G}(r)/\partial r$ represents the deterministic thermodynamic driving force derived from the free energy gradient. The second term $\xi(t)$ is a stochastic noise term reflecting the thermal fluctuations, modeled as Gaussian white noise satisfying standard statistical properties
$$\langle \xi(t) \rangle = 0, \quad \langle \xi(t)\xi(t') \rangle = 2D \delta(t-t'),$$
where $D$ is the diffusion coefficient. According to the fluctuation dissipation theorem, $D$ is related to the friction coefficient and the ensemble temperature $T$ via the Einstein relation $D = T/\zeta$.

Alternatively, the stochastic dynamics can be formulated in terms of the time evolution of the probability distribution $P(r, t)$, which satisfies the overdamped Fokker-Planck equation \cite{Kampen}
\begin{equation}
\frac{\partial P(r, t)}{\partial t} = \frac{\partial}{\partial r} \left( \frac{\partial \mathcal{G}(r)}{\partial r} P(r, t) \right) + D \frac{\partial^2 P(r, t)}{\partial r^2}=-\frac{\partial \mathcal{J}\left(r,t\right)}{\partial r}. \label{FPE}
\end{equation}
Here $\mathcal{J}(r,t)$ is themacroscopic probability flux 
\begin{equation}
    \mathcal{J}\left(r, t\right)=-\frac{\partial \mathcal{G}(r)}{\partial r} P\left(r, t\right)-D \frac{\partial P\left(r, t\right)}{\partial r}. \label{flux}
\end{equation}
These governing equations reveal that the dynamical behavior of the black hole is strictly dictated by the topography of the generalized free energy landscape $\mathcal{G}(r)$. The local extrema of this landscape define the stability conditions for different black hole equilibrium phases. To elucidate the mechanisms of the phase transition, we numerically solve Eq.\eqref{FPE} starting from a deterministic initial state (Dirac $\delta$ distribution). Taking $T=0.785T_c$ as a representative case (indicated by the right vertical dashed line in Fig.\ref{swallowtail}), we specifically focus on two critical scenarios, i.e., the stochastic evolution from a metastable potential well and the relaxation dynamics from an unstable local maximum.

\subsection{Dynamics from the Metastable State}

We first consider the scenario where the black hole is initialized at the metastable local minimum $r_s$. In this regime, the system is confined by a potential barrier and can only initiate a phase transition via stochastic thermal fluctuations. The characteristic timescale for this barrier crossing event is quantified by the Kramers escape time $\tau_K$\cite{Kramers:1940}, which is scaled exponentially with the ratio of the barrier height $\Delta \mathcal{G}$ to the noise strength $D$
\begin{equation}
\tau_K \propto \exp\left(\frac{\Delta \mathcal{G}}{D}\right). \label{Kramers}
\end{equation}
This exponential dependence implies that for a small diffusion coefficient $D$, the phase transition probability becomes negligible, effectively trapping the system in the metastable state.

\begin{figure}[!h]
\centering
\subfigure[$\Delta \mathcal{G} \gg D$ (Kinetic Trapping)]{
\includegraphics[width=0.45\textwidth]{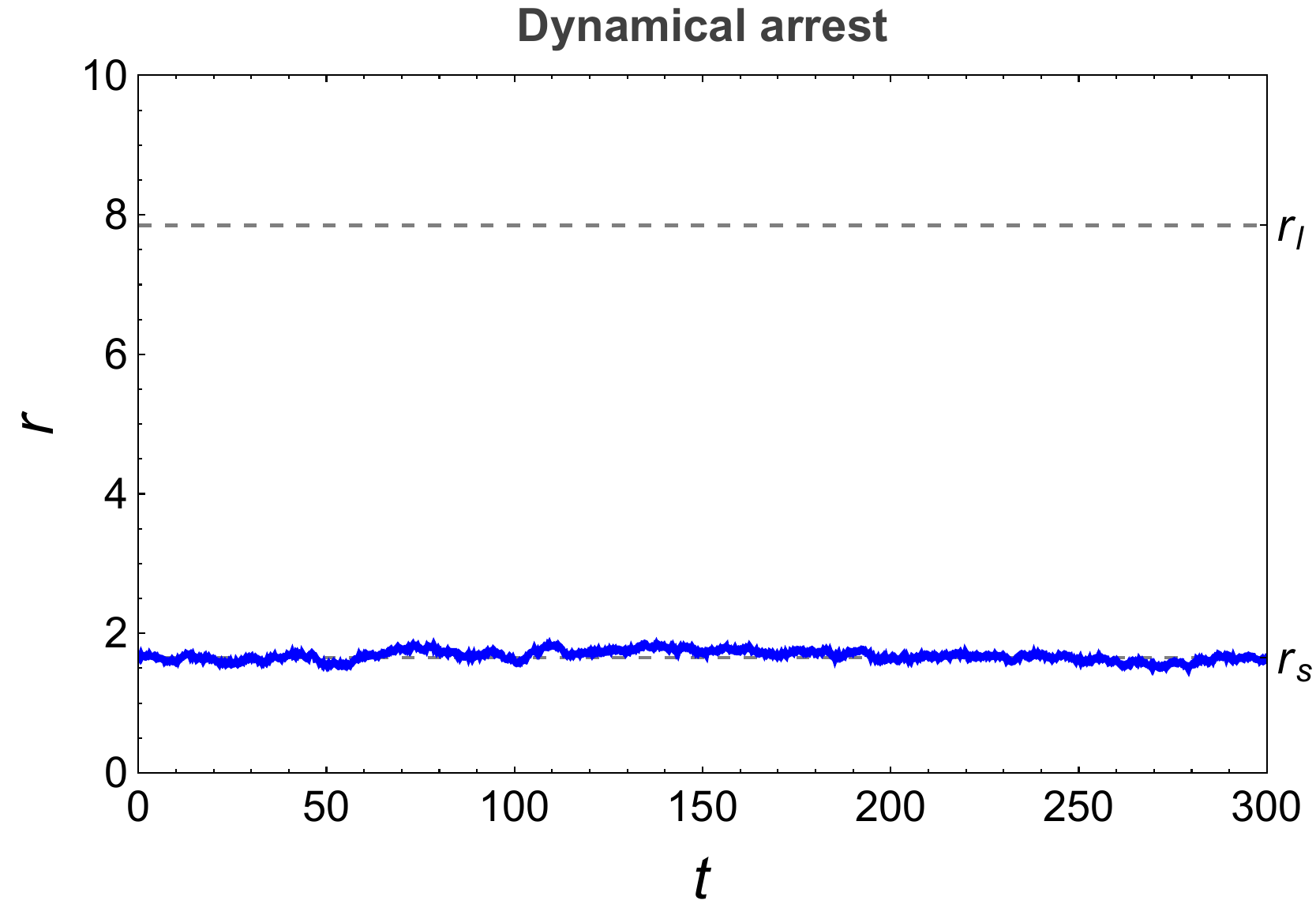}
}
\quad
\subfigure[$\Delta \mathcal{G} \lesssim D$ (Thermally Activated Escape)]{
\includegraphics[width=0.45\textwidth]{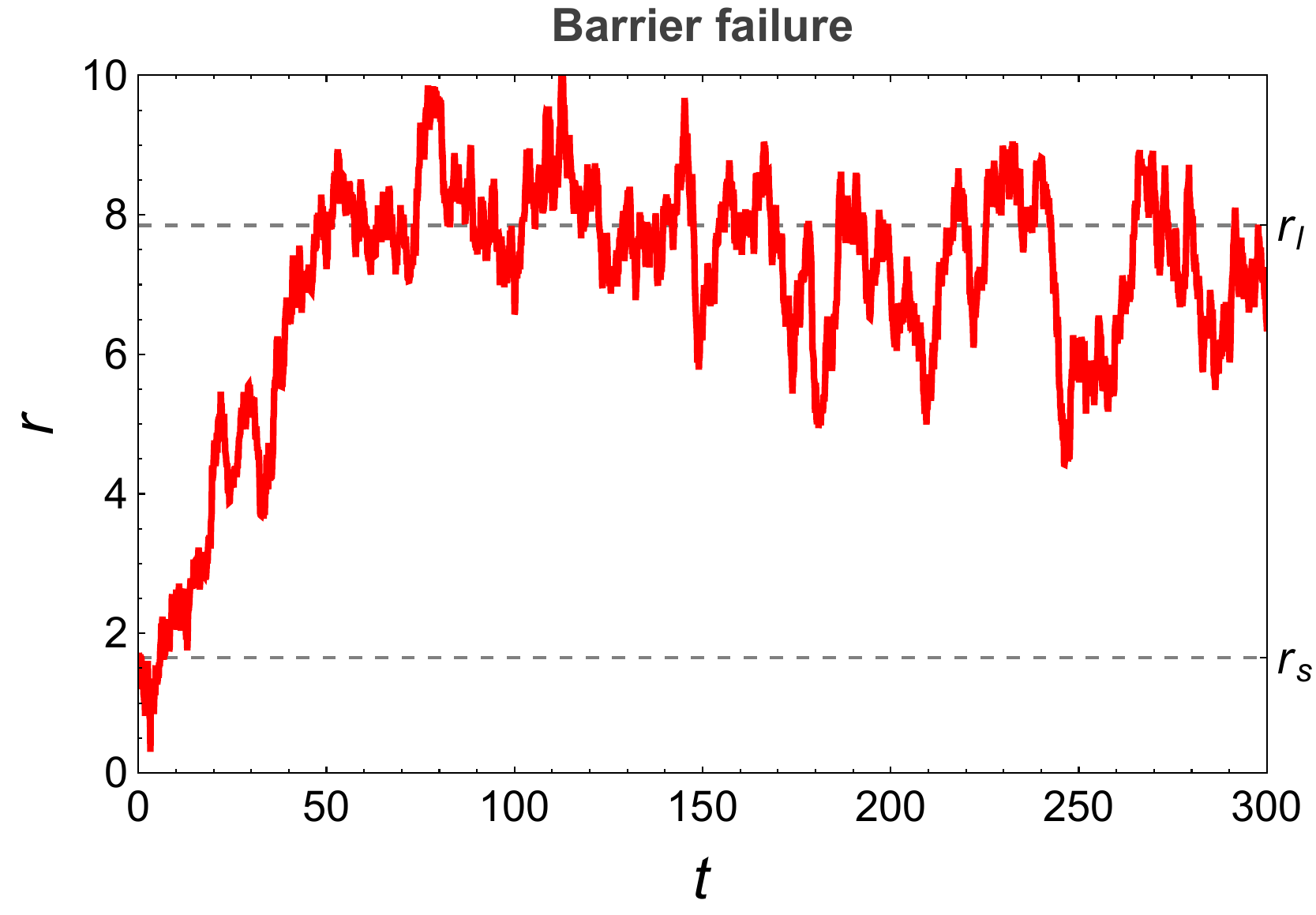}
}
\caption{Stochastic trajectories of the horizon radius $r$. The grey dashed lines indicate the horizon radii corresponding to the metastable small black hole ($r_s$) and the stable large black hole ($r_l$). (a) Under low diffusion, the system oscillates stochastically but remains confined within the metastable well. (b) Under high diffusion, the system accumulates sufficient energy to cross the barrier and relaxes into the globally stable well.}\label{langr}
\end{figure}

To visualize these distinct regimes, we depict Langevin simulation trajectories starting from the small black hole state at $T=0.785T_c$ in Fig.\ref{langr}.
In the low diffusion regime ($\Delta \mathcal{G} \gg D$), the Kramers escape time becomes exponentially large. Consequently, as shown in Fig.\ref{langr}(a), the black hole (blue trajectory) remains kinetically trapped, merely oscillating around the local minimum. Conversely, when diffusion is sufficient ($\Delta \mathcal{G} \lesssim D$), phase transition becomes statistically favorable. As depicted in Fig.\ref{langr}(b), the system (red trajectory) accumulates enough energy to surmount the potential peak and relaxes into the globally stable large black hole state.

\subsubsection{Kinetic Trapping and Spinodal Instability}

In the kinetic trapping regime, although the state does not correspond to the global most stable state and the phase transition is thermodynamically favorable, it is kinetically suppressed by the potential barrier. To analyze the relaxation process within the metastable well, we employ a linear approximation of the stochastic dynamics.

\begin{figure}[!h]
\centering
\subfigure[Initial State]{
\includegraphics[width=0.45\textwidth]{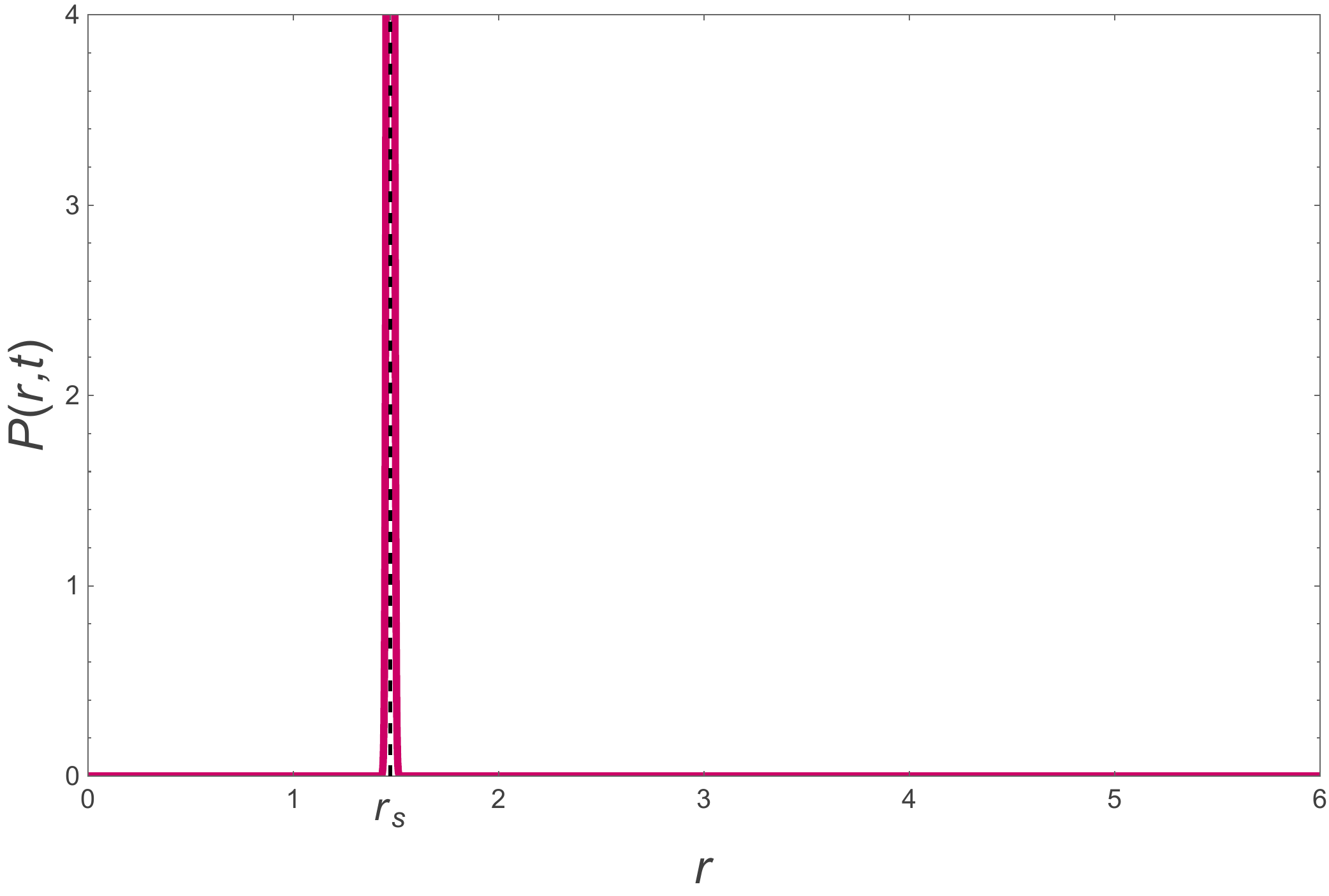}
}
\quad
\subfigure[Stationary State]{
\includegraphics[width=0.45\textwidth]{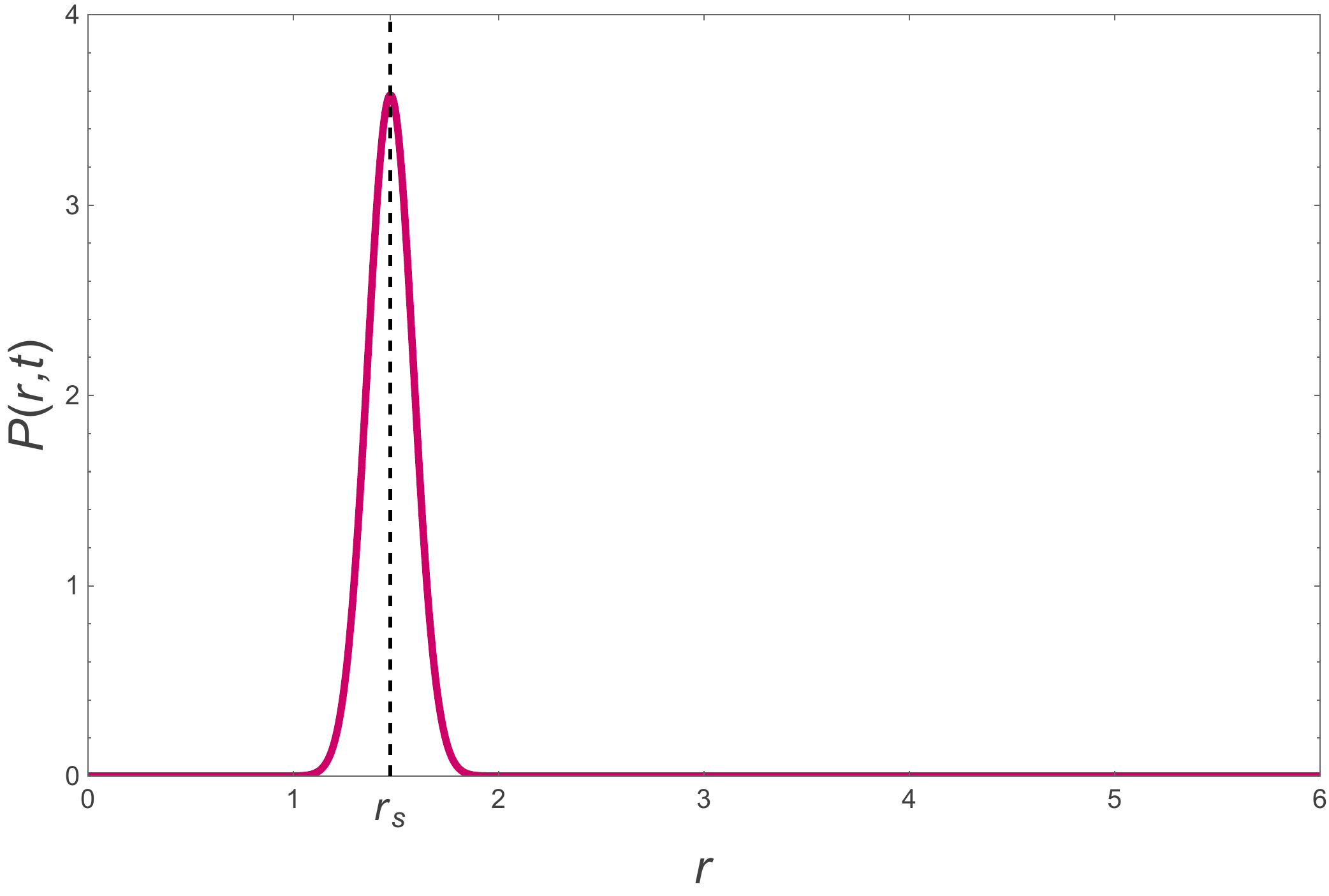}
}
\caption{Temporal evolution of the probability distribution $P(r, t)$ governed by the Fokker-Planck equation. (a) The initial distribution, highly localized at the metastable state. (b) The final stationary distribution where the system has fully relaxed into the globally stable state driven by thermal fluctuations. }\label{left}
\end{figure}

\begin{figure}[!ht]
\centering
\includegraphics[width=8cm]{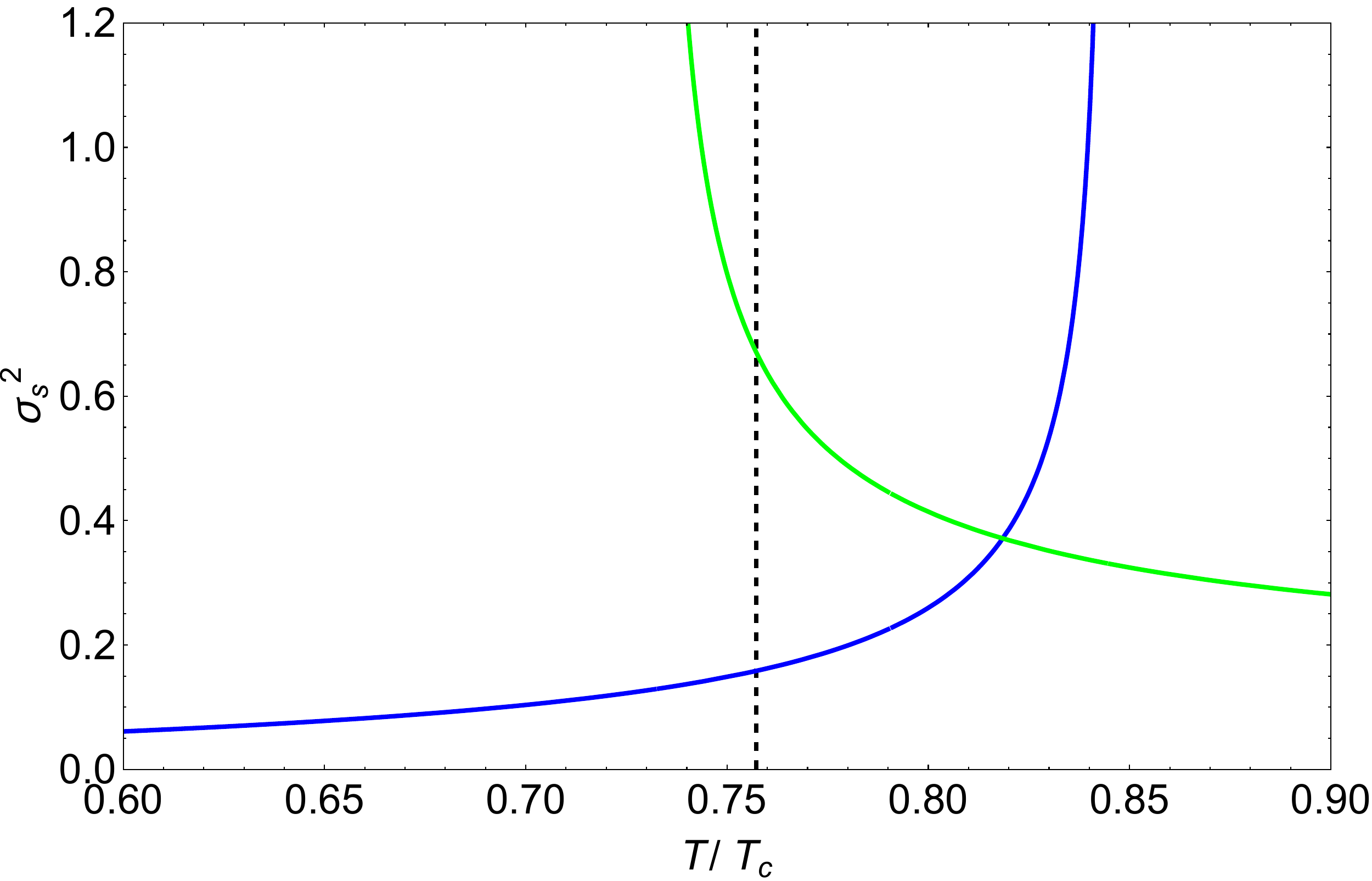}	
\caption{Stationary variance $\sigma_s^2$ versus reduced temperature $T/T_c$ for the SBH (blue) and LBH (green) phases. The vertical dashed line marks the coexistence state of the first-order transition, precisely corresponding to the intersection of stable branches in the swallowtail diagram (Fig.\ref{swallowtail}). }\label{sigma}
\end{figure}

We consider small fluctuations around the metastable local minimum $r_s$. Expanding the generalized free energy $\mathcal{G}(r)$ to second order in the deviation $x = r - r_s$, we obtain
\begin{equation}
\mathcal{G}(r) \approx \mathcal{G}(r_s) + \frac{1}{2} \alpha x^2,\label{sig}
\end{equation}
where $\alpha \equiv \mathcal{G}''(r_s) > 0$. Substituting this harmonic approximation into the general Fokker-Planck equation Eq.(\ref{FPE}), the dynamics reduces to the linear Fokker-Planck equation described as
\begin{equation}
\frac{\partial P(x, t)}{\partial t} =\frac{\partial ( \alpha x P)}{\partial x}  + D \frac{\partial^2 P}{\partial x^2}. \label{linearFP}
\end{equation}
Here, the drift term $\alpha x$ represents the linear restoring force that pulls the system back to equilibrium against fluctuations. For a system initialized at the potential bottom, represented by the Dirac delta distribution $P(x, 0) = \delta(x)$, this equation admits an exact time dependent Gaussian solution
\begin{equation}
P(x, t) = \sqrt{\frac{\alpha}{2\pi D (1 - e^{-2\alpha t})}} \exp\left( -\frac{\alpha x^2}{2D (1 - e^{-2\alpha t})} \right). \label{GaussSol}
\end{equation}
The width of this distribution is characterized by the time dependent variance $\sigma^2(t)$, which evolves as
\begin{equation}
\sigma^2(t) = \langle x^2(t) \rangle = \frac{D}{\alpha} (1 - e^{-2\alpha t}). \label{VarTime}
\end{equation}
The temporal evolution of this probability distribution is visualized in Fig.\ref{left}.
Starting from the perfectly localized state Fig.\ref{left}(a), the distribution undergoes Gaussian broadening driven by thermal diffusion. Over sufficiently long timescales, the exponential term in Eq.\eqref{VarTime} vanishes, and the packet saturates into a stationary symmetric Gaussian profile as in Fig.\ref{left}(b) with a finite stationary variance
\begin{equation}
\sigma_s^2 = \lim_{t \to \infty} \sigma^2(t) = \frac{D}{\mathcal{G}''(r_s)}.
\end{equation}
Physically, this stationary distribution represents a state of local equilibrium maintained by continuous thermal fluctuations around the free energy minimum. The magnitude of these fluctuations, explicitly quantified by the stationary variance $\sigma_s^2$, serves as a sensitive probe for the local stability of the black hole phase. The dependence of $\sigma_s^2$ on the heat bath temperature is illustrated in Fig.\ref{sigma}.
To explicitly demonstrate the connection between the macroscopic thermodynamic phase structure and the stochastic fluctuations, the horizontal axis of Fig.\ref{sigma} is strictly aligned with that of the isobaric swallowtail diagram in Fig.\ref{swallowtail}. Visually, the divergent vertical asymptotes of $\sigma_s^2$ perfectly coincide with the endpoints (cusps) of the swallowtail branches in Fig.\ref{swallowtail}. As the temperature approaches these spinodal limits, the metastable minimum and the unstable maximum merge, causing the local curvature of the free energy landscape to strictly vanish ($\mathcal{G}''(r_s) \to 0$). Consequently, the stationary variance diverges as
\begin{equation}
	\sigma_s^2 = \frac{k_B T}{\mathcal{G}''(r_s)} \to \infty.
\end{equation}
This divergence signifies the complete disappearance of the thermodynamic restoring force. The flattening of the potential well leaves the black hole infinitely susceptible to thermal noise, marking the ultimate breakdown of local mechanical stability.

\subsubsection{Phase Transition }

\begin{figure}[!h]
	\centering
	\subfigure[Initial state]{
		\includegraphics[width=4.5cm]{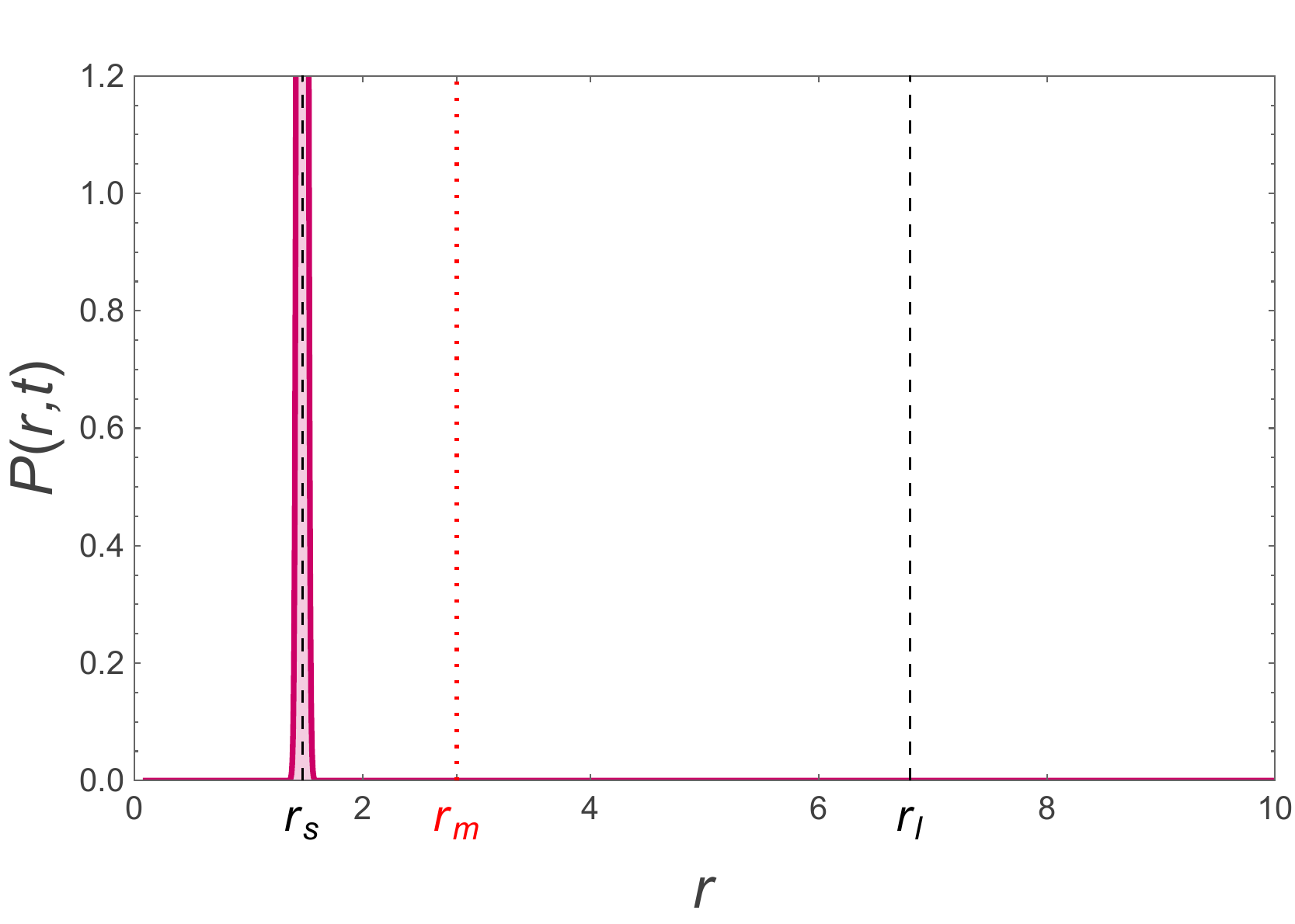}
	}
	\quad
	\subfigure[Local relaxation]{
		\includegraphics[width=4.5cm]{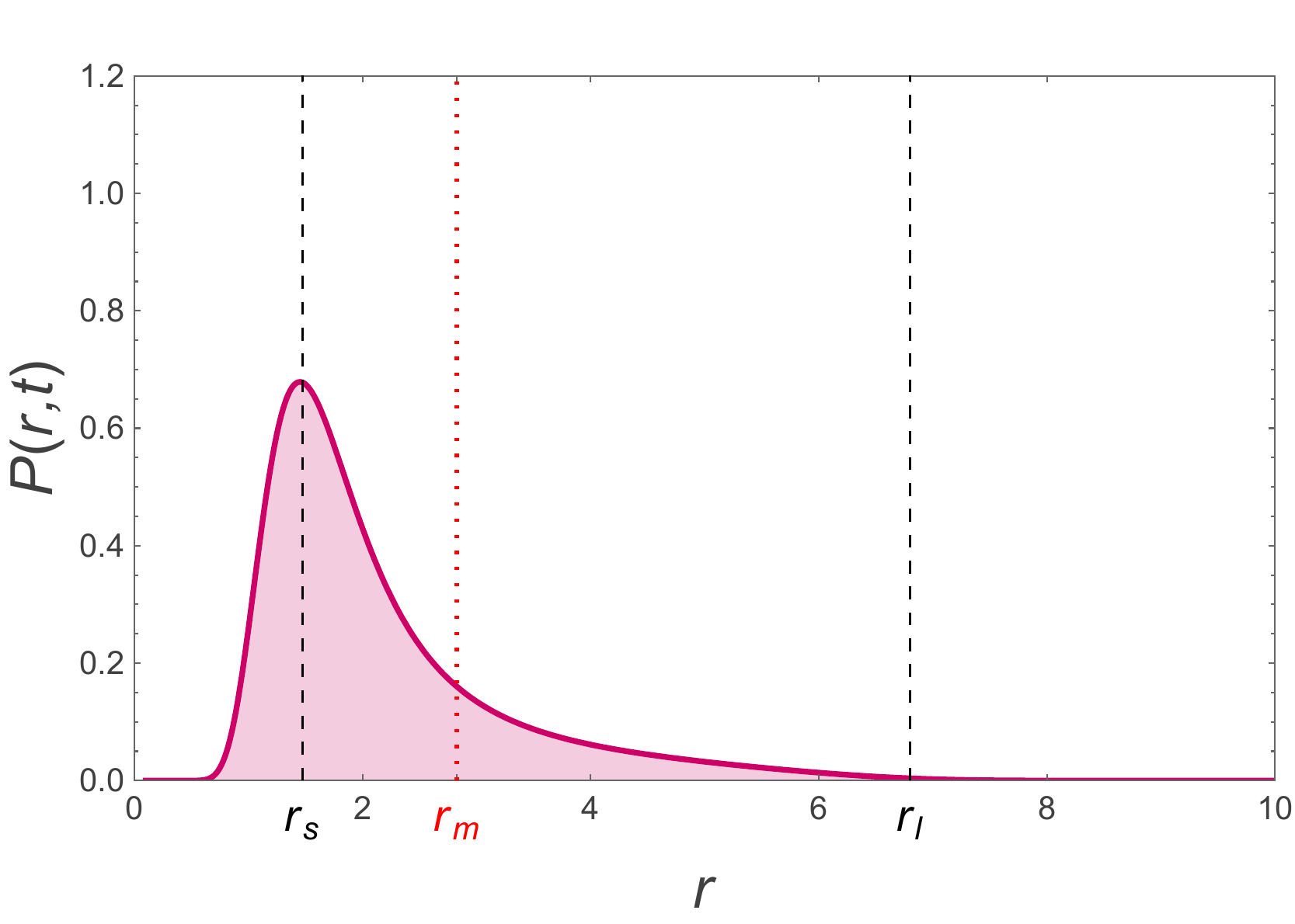}
	}
	\quad
	\subfigure[Onset of transition]{
		\includegraphics[width=4.5cm]{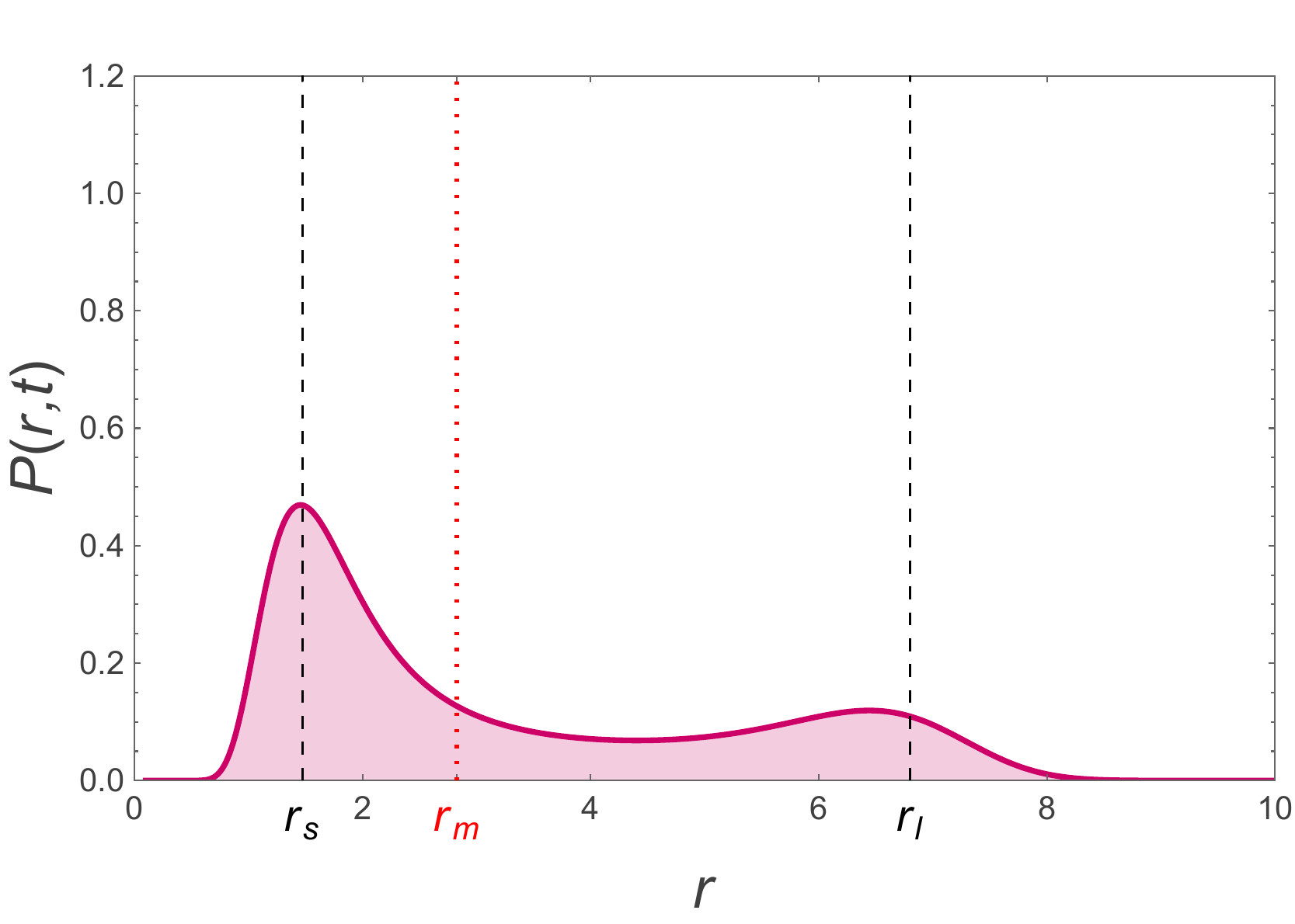}
	}
	\quad
	\subfigure[Bimodal transient]{
		\includegraphics[width=4.5cm]{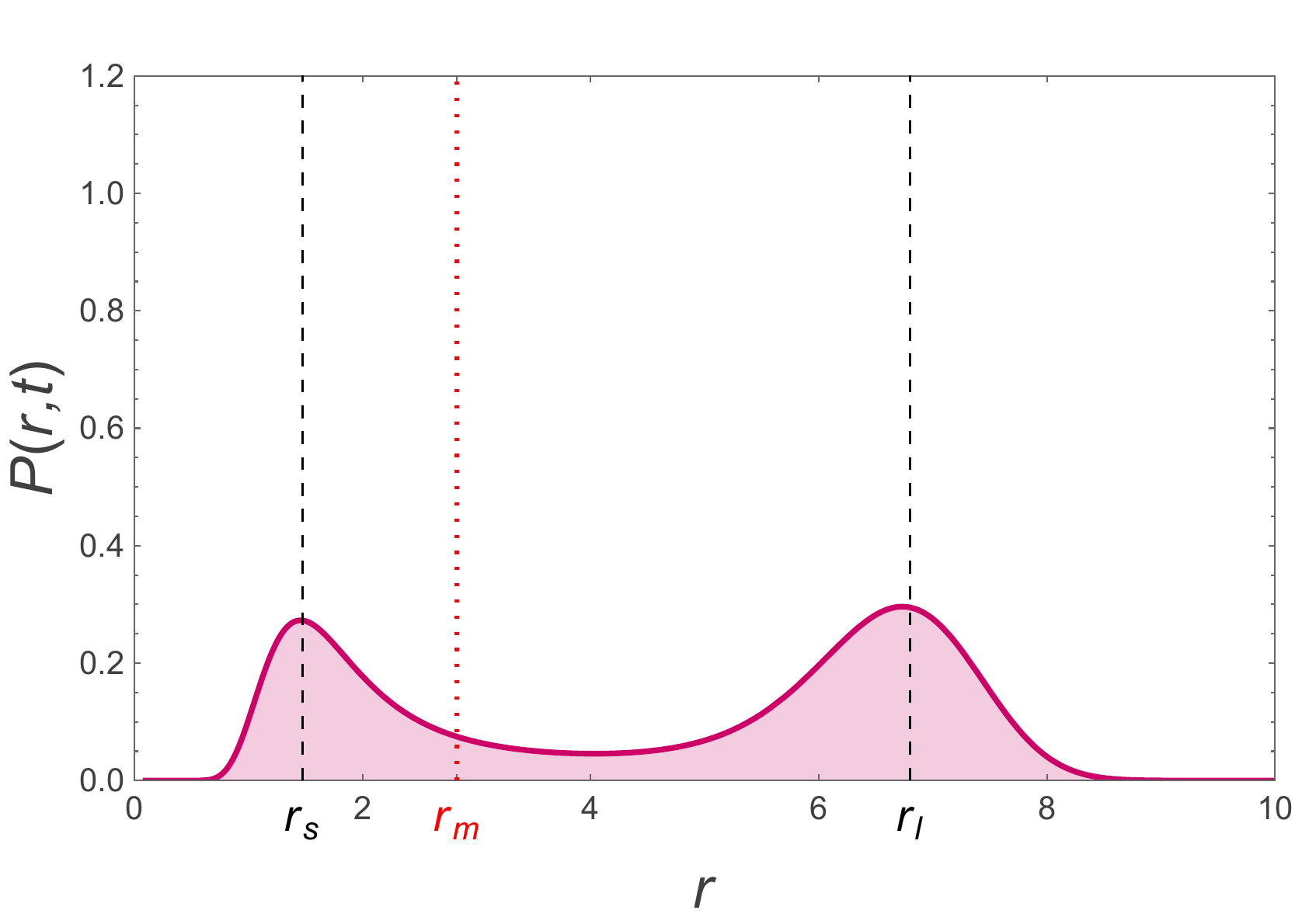}
	}
	\quad
	\subfigure[Approaching stationary]{
		\includegraphics[width=4.5cm]{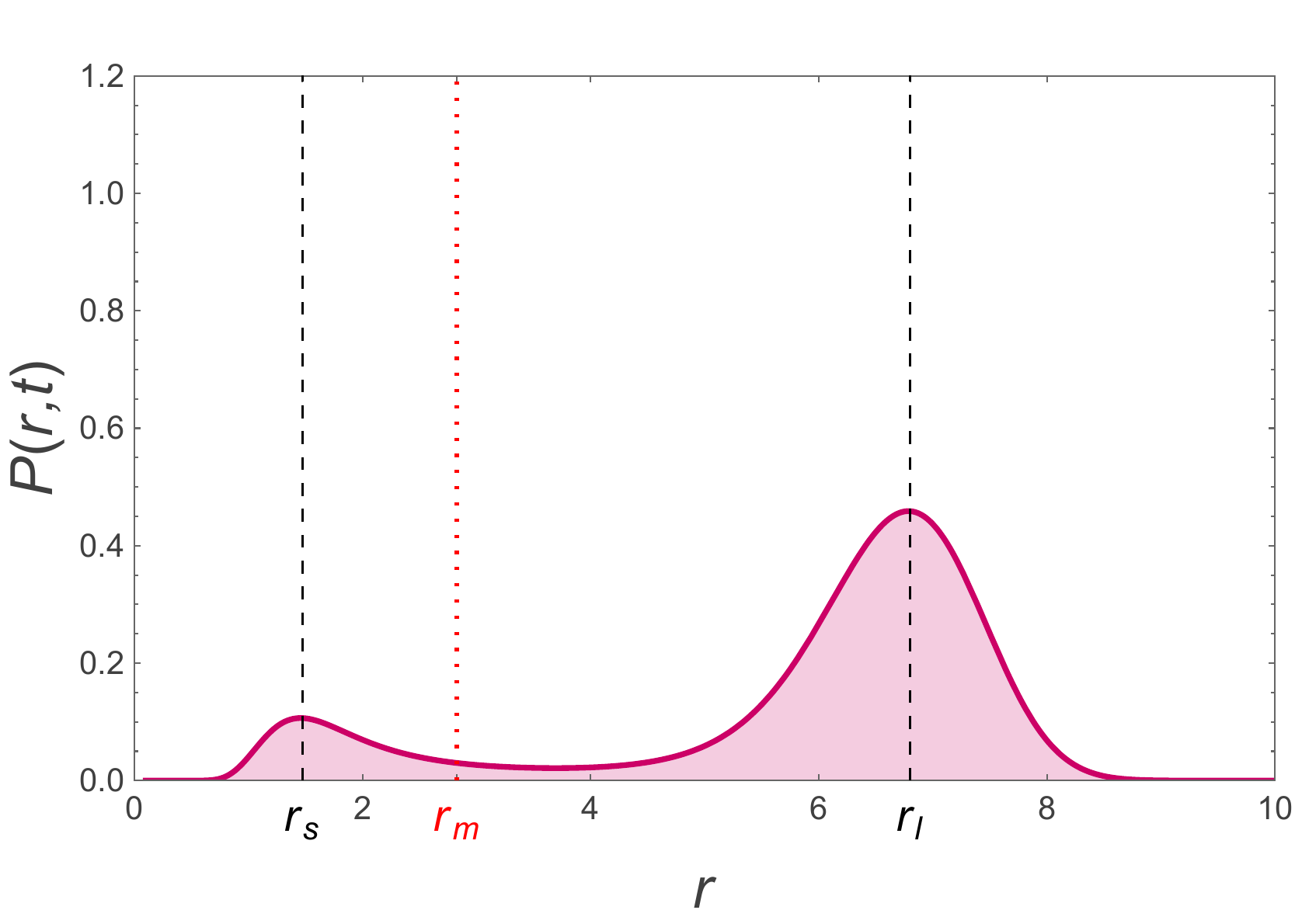}
	}
	\quad
	\subfigure[Stationary state]{
		\includegraphics[width=4.5cm]{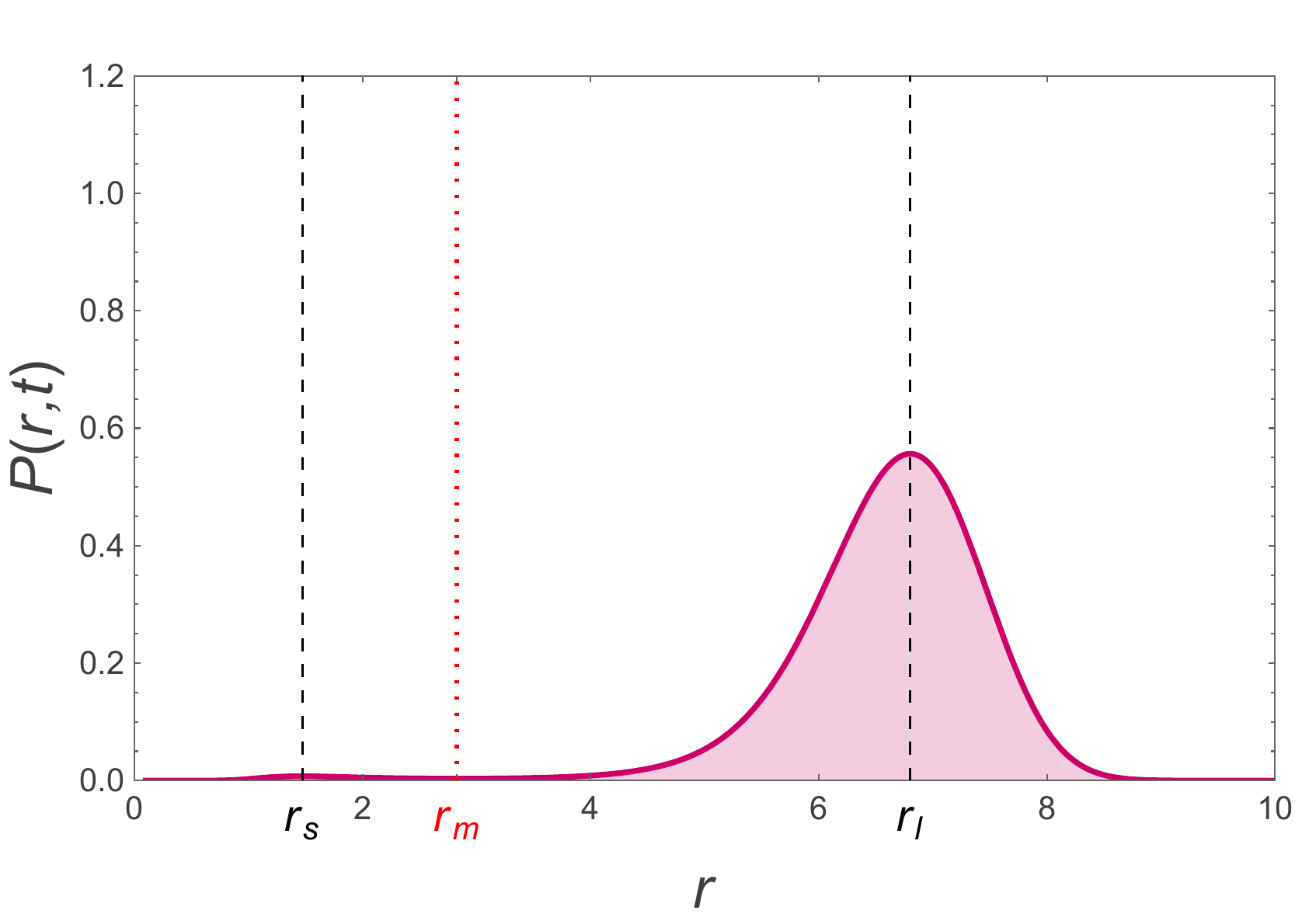}
	}
	\caption{Time evolution of the probability distribution $P(r_h, t)$ during the first order phase transition.}\label{escape}
\end{figure}

In the high diffusion regime where thermal fluctuations are comparable to the energy barrier ( $\Delta \mathcal{G} \lesssim D$ ), the system undergoes a dynamical phase transition. The full probabilistic evolution of this stochastic process is governed by the Fokker-Planck equation. To ensure the strict conservation of total probability during the entire dynamic evolution, we impose reflecting boundary conditions, mathematically formulated as zero flux Neumann boundary conditions. Therefore, the macroscopic probability flux $\mathcal{J}\left(r, t\right)$ in Eq.(\ref{flux}) is required to completely vanish at the physical boundaries, which is
\begin{equation}
\left.J(r, t)\right|_{r \rightarrow 0}=\left.J(r, t)\right|_{r \rightarrow \infty}=0 .\label{boundary}
\end{equation}
The numerical solution of this transition, from the metastable confinement to the globally stable configuration, is explicitly visualized in Fig.\ref{escape}. 
Initially, the probability distribution is prepared as a highly localized wave packet within the metastable potential well (Fig.\ref{escape}(a)). Driven by thermal noise, the system first undergoes a rapid local relaxation, naturally broadening within the metastable state (Fig.\ref{escape}(b)). As time proceeds, stochastic fluctuations drive a continuous probability flux across the potential barrier at $r_m$ (marked by the red dotted line), triggering the onset of the transition (Fig.\ref{escape}(c)). This continuous leakage leads to the emergence of a striking bimodal transient (Fig.\ref{escape}(d)), indicating that the system statistically explores both the metastable ($r_s$) and globally stable ($r_l$) basins simultaneously. Ultimately, the global thermodynamic driving force prevails, and the probability weight is progressively transferred (Fig.\ref{escape}(e)) until the system completely settles into the final stationary state (Fig.\ref{escape}(f)).

A critical feature of this final stationary state is that the probability distribution does not collapse into a singular Dirac $\delta$ function but rather stabilizes as a broad Gaussian profile. While a Dirac $\delta$ distribution would imply vanishing Shannon entropy (i.e., zero macroscopic uncertainty regarding the order parameter), the final state here represents a dynamic thermodynamic equilibrium. In this regime, the stationary variance $\sigma_l^2= k_B T / \mathcal{G}^{\prime \prime}\left(r_l\right)$ reflects the continuous balance between the deterministic restoring force and the stochastic thermal noise, indicating that even in the globally stable phase, the black hole maintains intrinsic geometric fluctuations.

\subsection{Dynamics from the Unstable Maximum}

\begin{figure}[!h]
\centering
\subfigure[Initial state]{
\includegraphics[width=4.5cm]{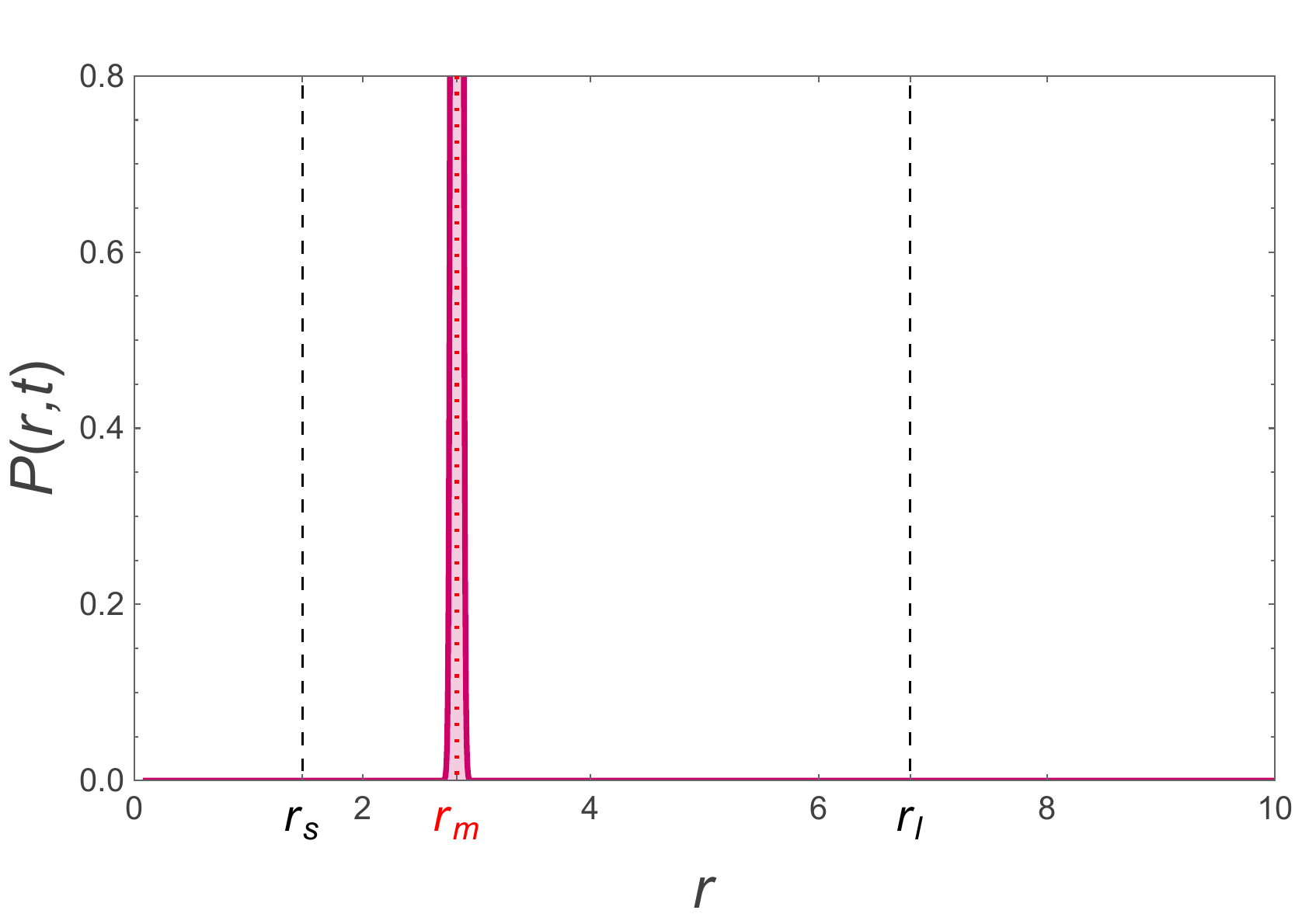}
}
\quad
\subfigure[Initial broadening]{
\includegraphics[width=4.5cm]{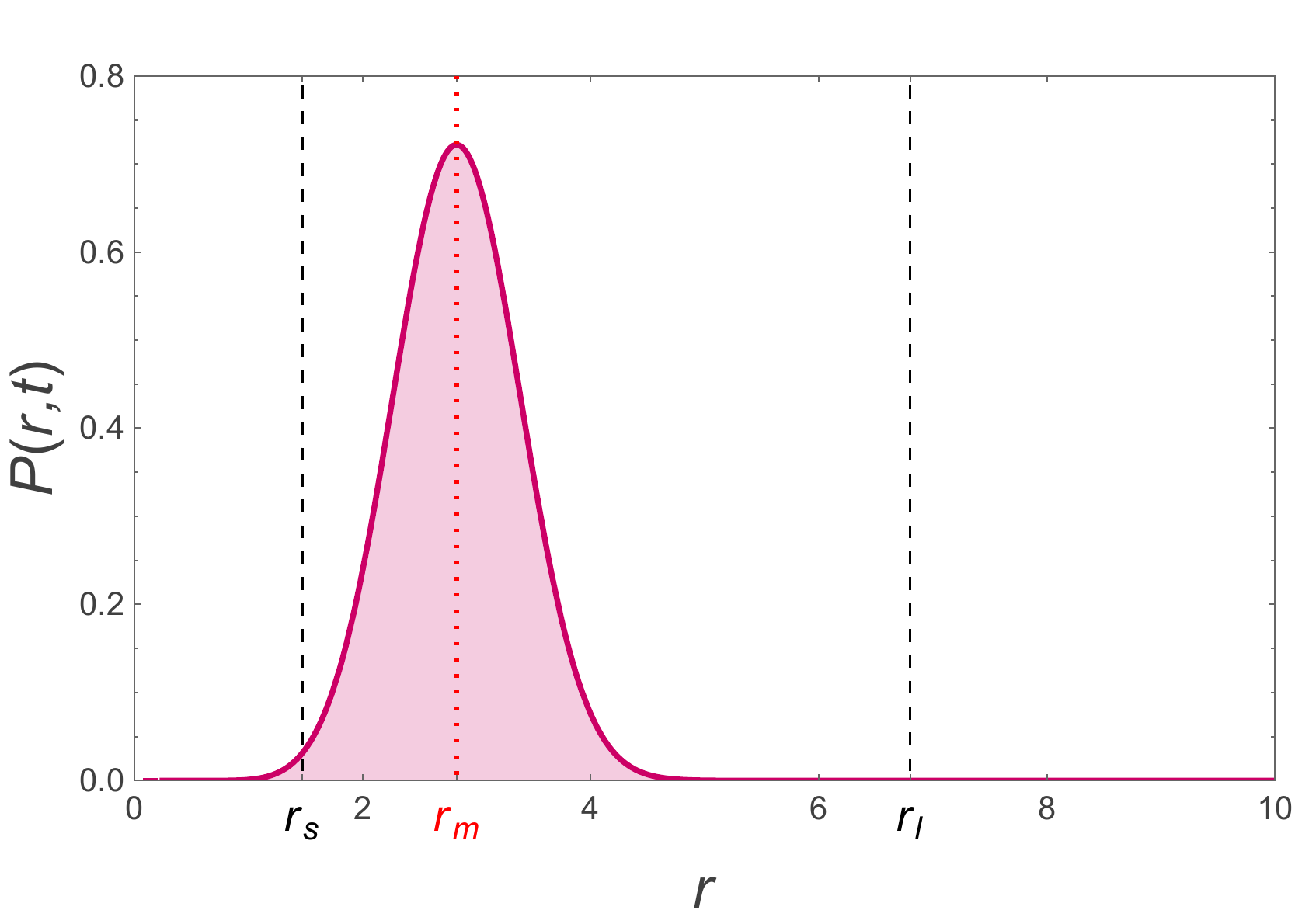}
}
\quad
\subfigure[Wave packet splitting]{
\includegraphics[width=4.5cm]{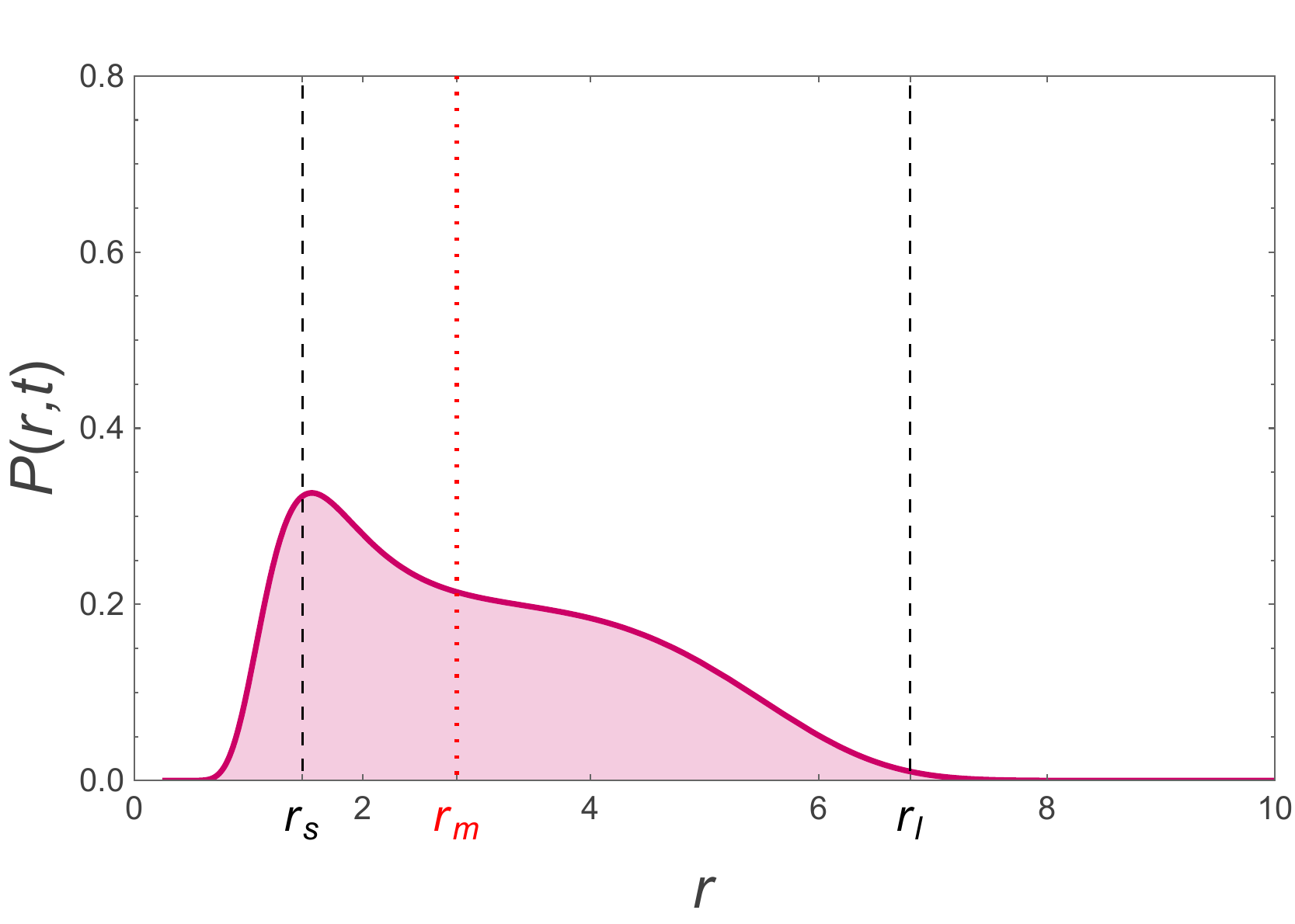}
}
\quad
\subfigure[Bimodal transient]{
\includegraphics[width=4.5cm]{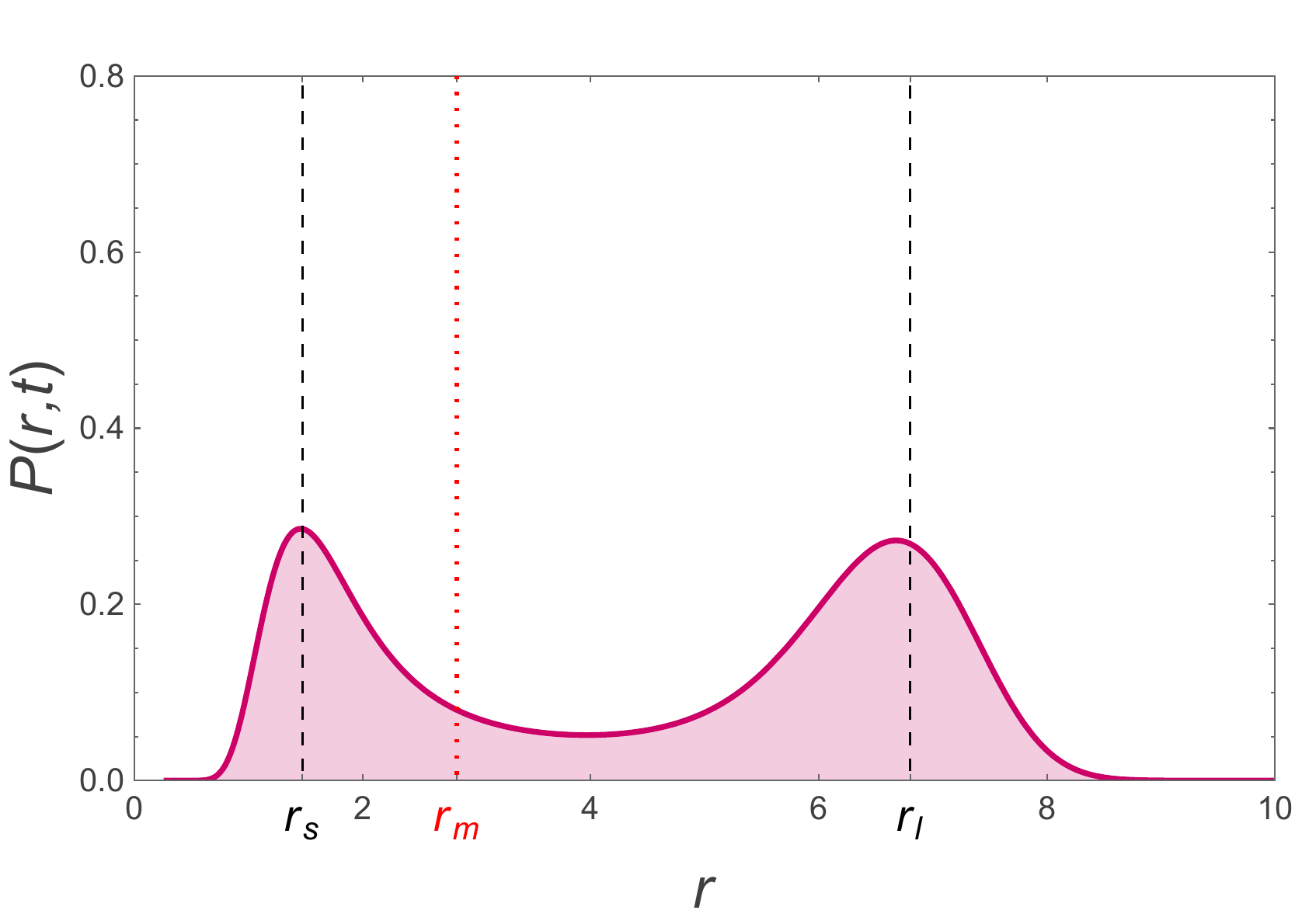}
}
\quad
\subfigure[Approaching stationary]{
\includegraphics[width=4.5cm]{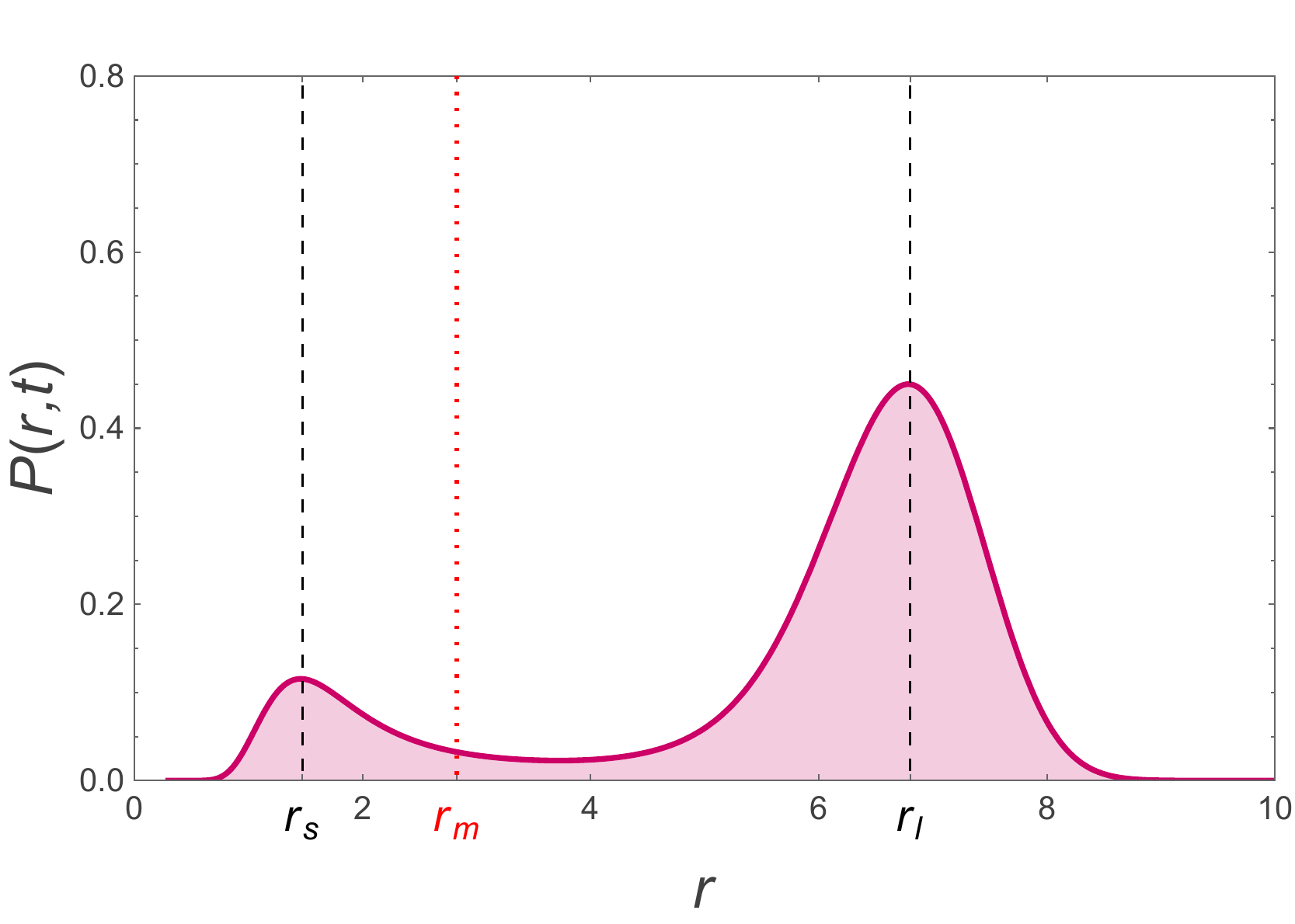}
}
\quad
\subfigure[Stationary state]{
\includegraphics[width=4.5cm]{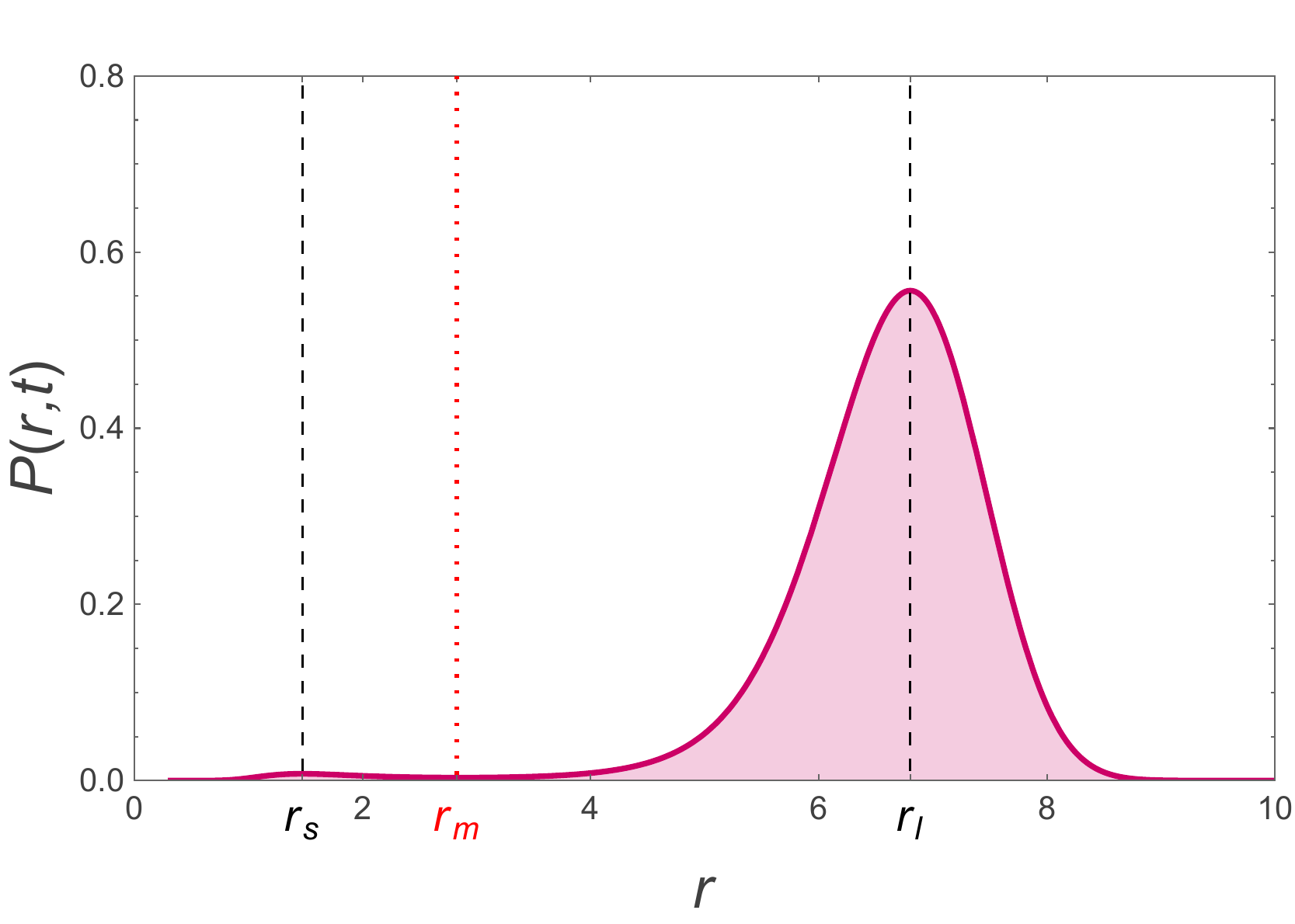}
}
\caption{Temporal evolution of the probability distribution $P(r_h, t)$ initialized at the unstable local maximum.}\label{localmiximum}
\end{figure}

We finally analyze the relaxation dynamics when the system is initialized at the unstable local maximum. Although the early stage evolution can still be described by the linearized Fokker-Planck formalism of Eq.(\ref{linearFP}), the physical nature of the landscape is fundamentally inverted. In stark contrast to the metastable minimum, the local curvature here is strictly negative ($\alpha=\mathcal{G}''(r_m) < 0$). This negative curvature transforms the drift term from a restoring force into a repulsive thermodynamic force. Consequently, the probability distribution stays Gaussian, but its width (variance) keeps increasing over time. Instead, by substituting the negative curvature into Eq.\eqref{VarTime}, the variance grows exponentially
\begin{equation} \sigma^2(t) = \frac{D}{|\alpha|} (e^{2|\alpha| t} - 1). \label{variance_unstable} 
\end{equation} 
Physically, this means the initial probability packet expands rapidly and symmetrically. This divergence indicates that the system is collapsing from the unstable peak, causing the probability to flow into the adjacent stable valleys. This linear approximation remains valid only within a restricted neighborhood around the potential maximum. Beyond this range, the non-linearity of the free energy landscape becomes dominant. The initial linear expansion process is generally considered to end when the standard deviation of the wave packet, $\sigma(t)$, reaches approximately half the distance to the nearest potential well. As visually evident in Fig.\ref{landscape}, the metastable state $r_s$ is spatially closer to the barrier peak $r_m$. Letting $L = |r_m - r_s|$ be this shortest characteristic distance, the boundary of the linear regime is strictly defined by the criterion $\sigma(t) \approx L/2$, or equivalently $\sigma^2(t) \approx L^2/4$. Substituting this critical threshold into the variance evolution equation, the characteristic timescale for the initial linear broadening is estimated as
	$$t \approx -\frac{1}{2\alpha} \ln\left( 1 - \frac{\alpha L^2}{4D} \right),$$
yielding a strictly positive time $t > 0$, given the negative curvature $\alpha < 0$ at the local maximum. To observe how the probability packet splits and flows into the stable basins, we numerically solve the full Fokker-Planck equation with zero flux Neumann boundary conditions in Eq.(\ref{boundary}). The results are plotted in Fig.\ref{localmiximum}.
The numerical simulation reveals that the relaxation process visually unfolds through several distinct stages. As depicted in Fig. \ref{localmiximum}(a), the system is initially prepared as a highly localized $\delta$ like peak at the barrier top. Here, the negative curvature acts as a repulsive force, driving the probability packet to undergo the initial symmetric broadening shown in Fig. \ref{localmiximum}(b). As the distribution expands beyond the linear regime, a significant asymmetry emerges, triggering the wave packet splitting illustrated in Fig. \ref{localmiximum}(c). Although the global minimum (right well, $r_l$ ) is thermodynamically more favorable, the probability flux preferentially flows into the metastable basin (left well, $r_s$ ). This kinetically driven process results in the bimodal transient distribution clearly visible in Fig. \ref{localmiximum}(d). Such a preferential flow occurs because the metastable well is spatially closer to the barrier peak, as previously shown in Fig. \ref{landscape}, causing the system to be temporarily trapped. Ultimately, driven by the global thermodynamic gradient, the probability weight continuously transfers across the landscape. Fig. \ref{localmiximum}(e) captures the system approaching stationary during this transfer, until it eventually relaxes into the true global stationary state presented in Fig. \ref{localmiximum}(f).

\section{Entropy and Irreversibility}\label{IV}

The Fokker-Planck equation yields a comprehensive probabilistic description of the black hole's evolution. Crucially, the distribution $P(r, t)$ quantifies the statistical uncertainty of the macroscopic order parameter (the horizon radius) as distinct from the black hole's internal microscopic degrees of freedom. To understand why the phase transition proceeds in a specific direction and to quantify the thermodynamic cost, we extend our analysis to the concepts of entropy and irreversibility. In the framework of stochastic thermodynamics, the phase transition is treated not merely as a geometric shift in the free energy landscape, but as a non-equilibrium process driven by the dissipation of information and energy.

To quantify the information theoretic properties and the degree of irreversibility of the stochastic dynamics, we utilize two key thermodynamic indicators. First, the Shannon Entropy measures the uncertainty associated with the macroscopic probability distribution $P(r, t)$
\begin{equation}
S(t) = - \int P(r, t) \ln P(r, t)  dr_h. \label{shannon}
\end{equation}
Second, the entropy production rate $\Pi(t)$ quantifies the thermodynamic cost and irreversibility of the phase transition. It measures how far the system deviates from equilibrium during its evolution. In the Fokker-Planck framework, $\Pi(t)$ is defined as the integral of the squared probability flux
\begin{equation}
\Pi(t) = \int \frac{\mathcal{J}^2(r, t)}{D P(r, t)}dr \geq 0, \label{epr}
\end{equation}
where $\mathcal{J}(r, t)$ is the probability current in Eq.(\ref{flux}). A non zero $\Pi(t)$ indicates that the system is out of equilibrium and is evolving towards a stable state.

\begin{figure}[!h]
\centering
\subfigure[Shannon Entropy]{
\includegraphics[width=0.45\textwidth]{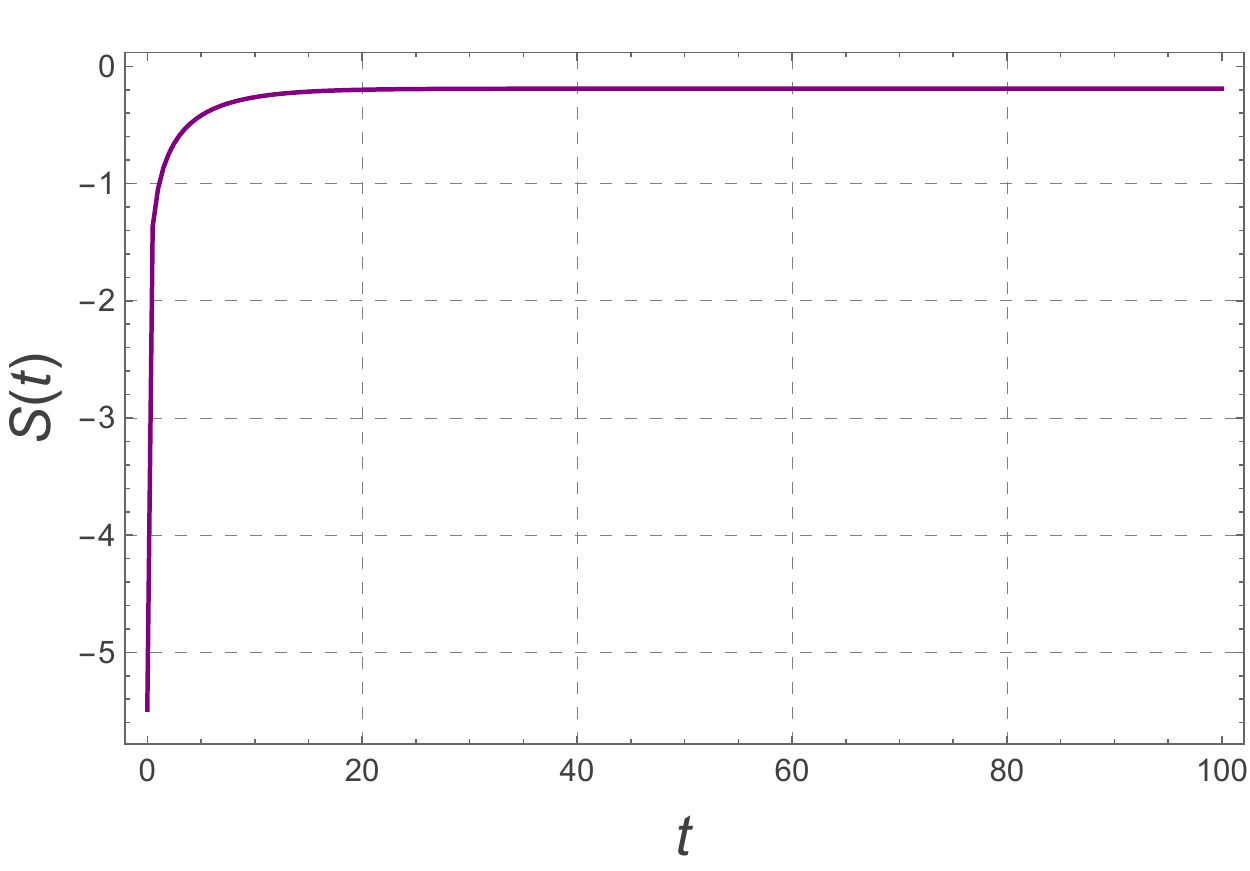}
}
\quad
\subfigure[Entropy Production Rate]{
\includegraphics[width=0.45\textwidth]{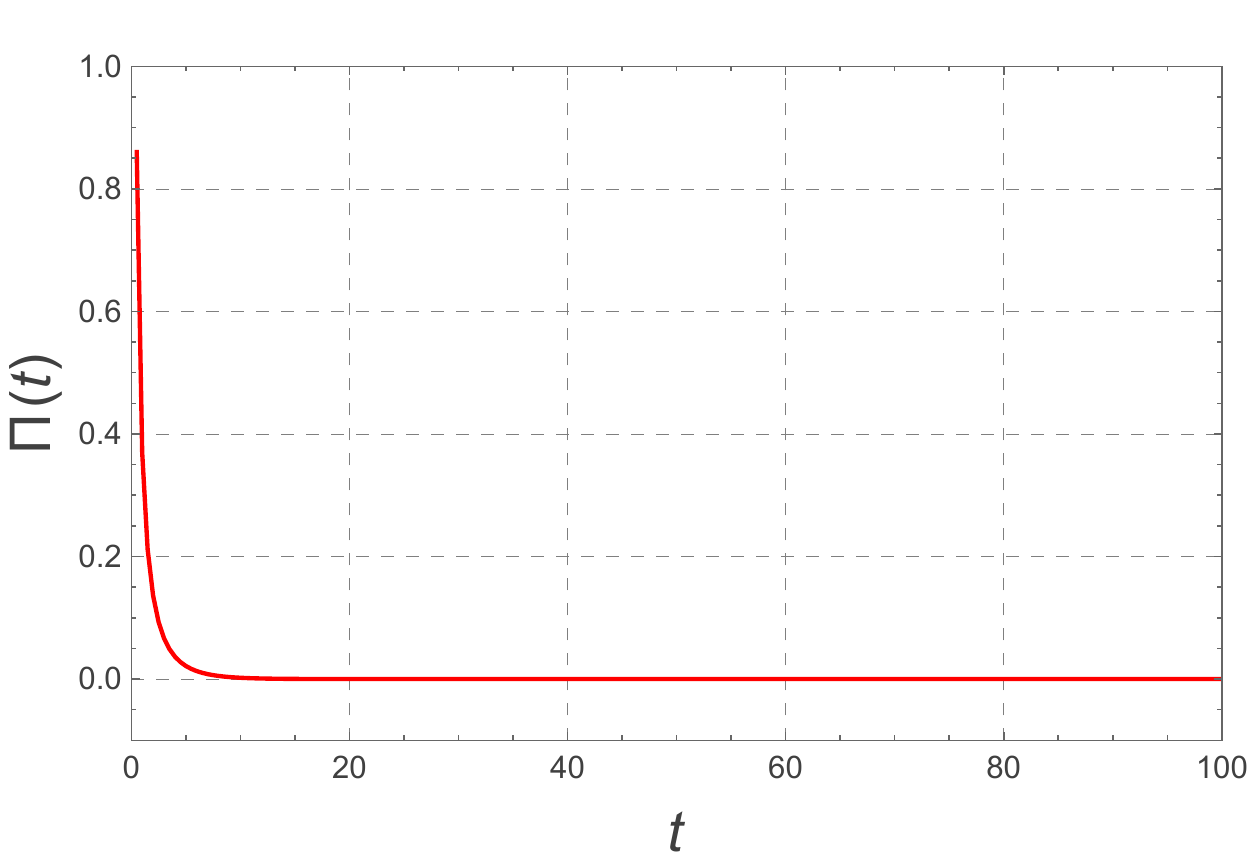}
}
\caption{Entropic evolution in the kinetic trapping regime. The system rapidly relaxes to a local equilibrium state.}\label{entropy_trapping}
\end{figure}

In the regime of low diffusion, the system is kinetically trapped within the metastable potential well. The entropic evolution is displayed in Fig.\ref{entropy_trapping}.
The plateauing of $S(t)$ confirms that the probability distribution has relaxed into a stationary Gaussian profile, marking the onset of local equilibrium. The black hole's fluctuations are now confined within the potential well. Correspondingly, as shown in Fig.\ref{entropy_trapping}(b), the entropy production rate $\Pi(t)$ decays to zero. This signifies that the net probability flow has stopped. Physically, thermal noise is now balanced by the potential force, resulting in zero dissipation despite the system being in a metastable state.

\begin{figure}[!h]
\centering
\subfigure[Shannon Entropy]{
\includegraphics[width=0.45\textwidth]{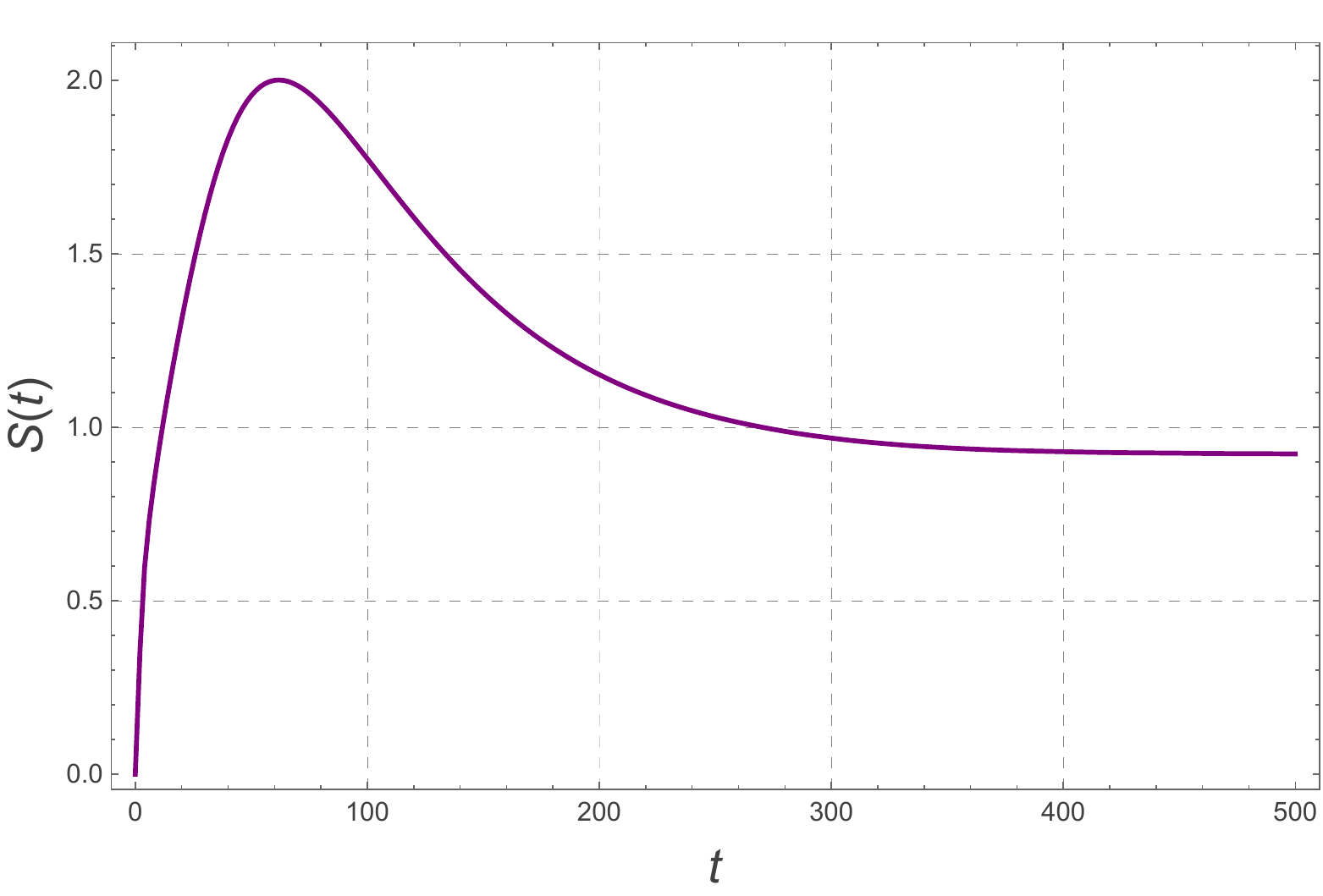}
}
\quad
\subfigure[Entropy Production Rate]{
\includegraphics[width=0.45\textwidth]{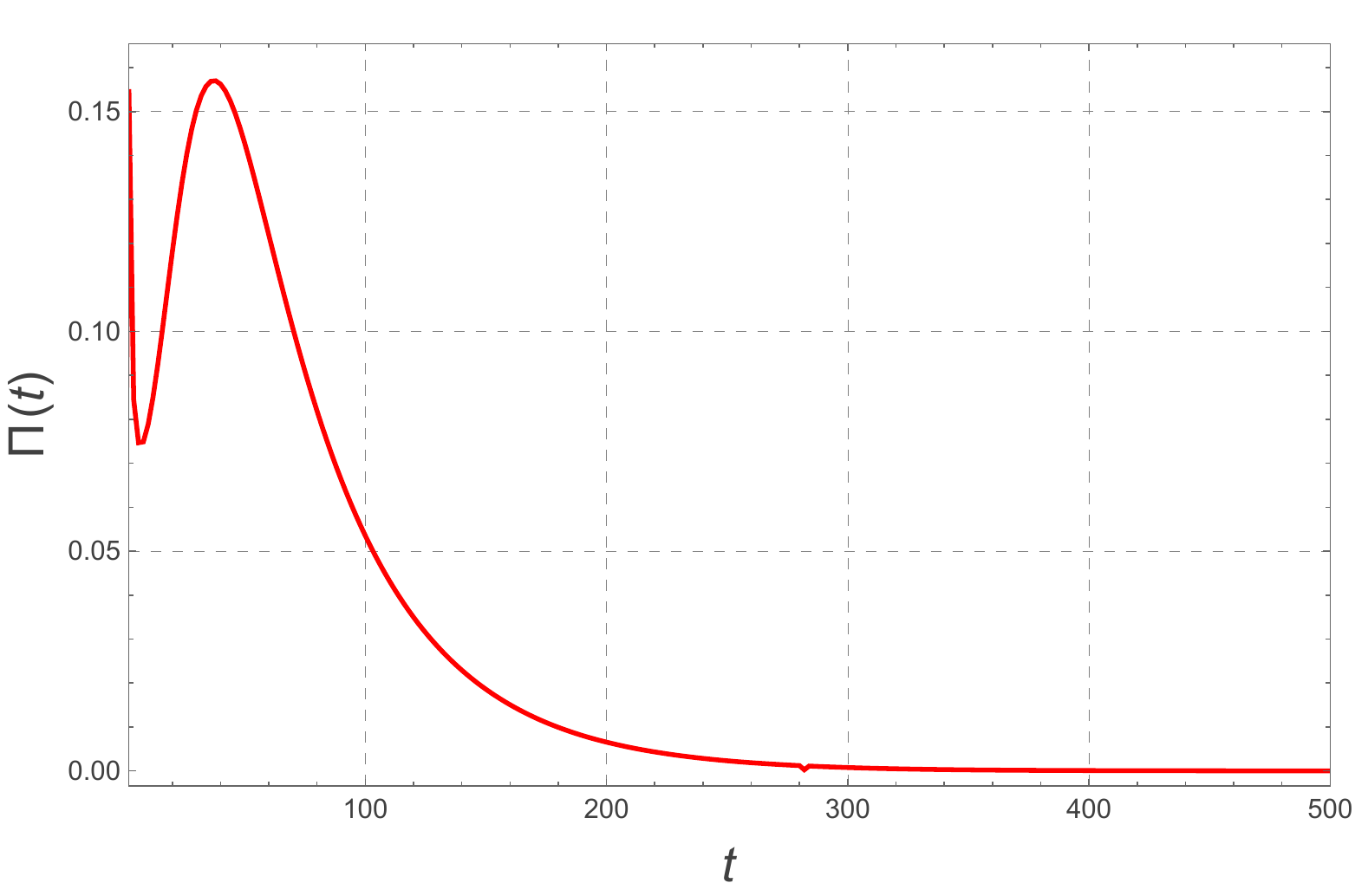}
}
\caption{Entropic evolution during a thermally activated phase transition.}\label{entropy_transition}
\end{figure}

With sufficient diffusion, the system overcomes the barrier and undergoes a phase transition. This process is captured in Fig.\ref{entropy_transition}.
As shown in Fig.\ref{entropy_transition}(a), $S(t)$ peaks as the system crosses the potential barrier, reflecting the moment of maximum uncertainty. It then stabilizes at a constant, non zero value, as thermal fluctuations maintain a finite distribution width even in the stable well. In Fig.\ref{entropy_transition}(b), the entropy production rate initially decreases due to relaxation within the metastable well. However, this decay is interrupted. As the system escapes the well and surmounts the potential barrier, $\Pi(t)$ rises again to form a prominent peak. This peak marks the critical moment of the phase transition, representing the intense thermodynamic dissipation required to drive the black hole from the metastable state to the stable phase.

\begin{figure}[!h]
\centering
\subfigure[Shannon Entropy]{
\includegraphics[width=0.45\textwidth]{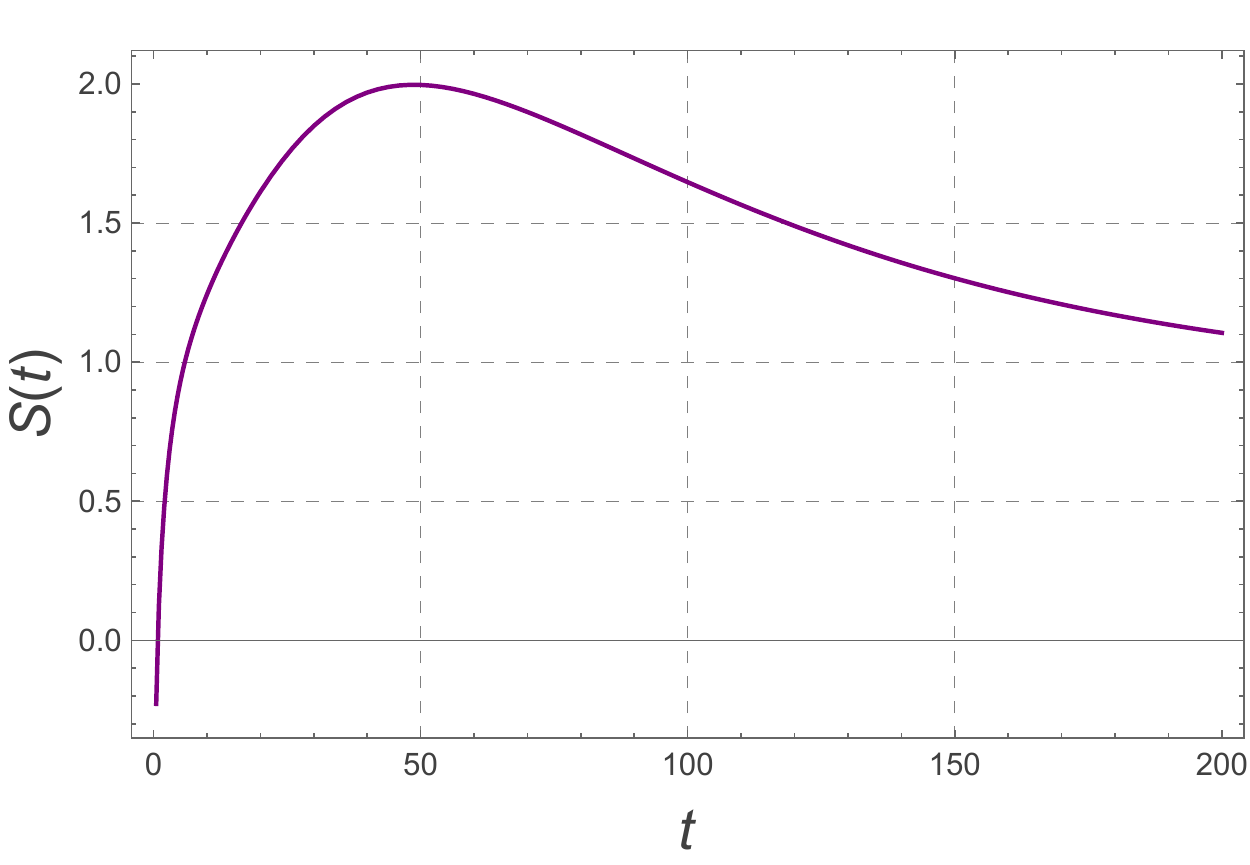}
}
\quad
\subfigure[Entropy Production Rate]{
\includegraphics[width=0.45\textwidth]{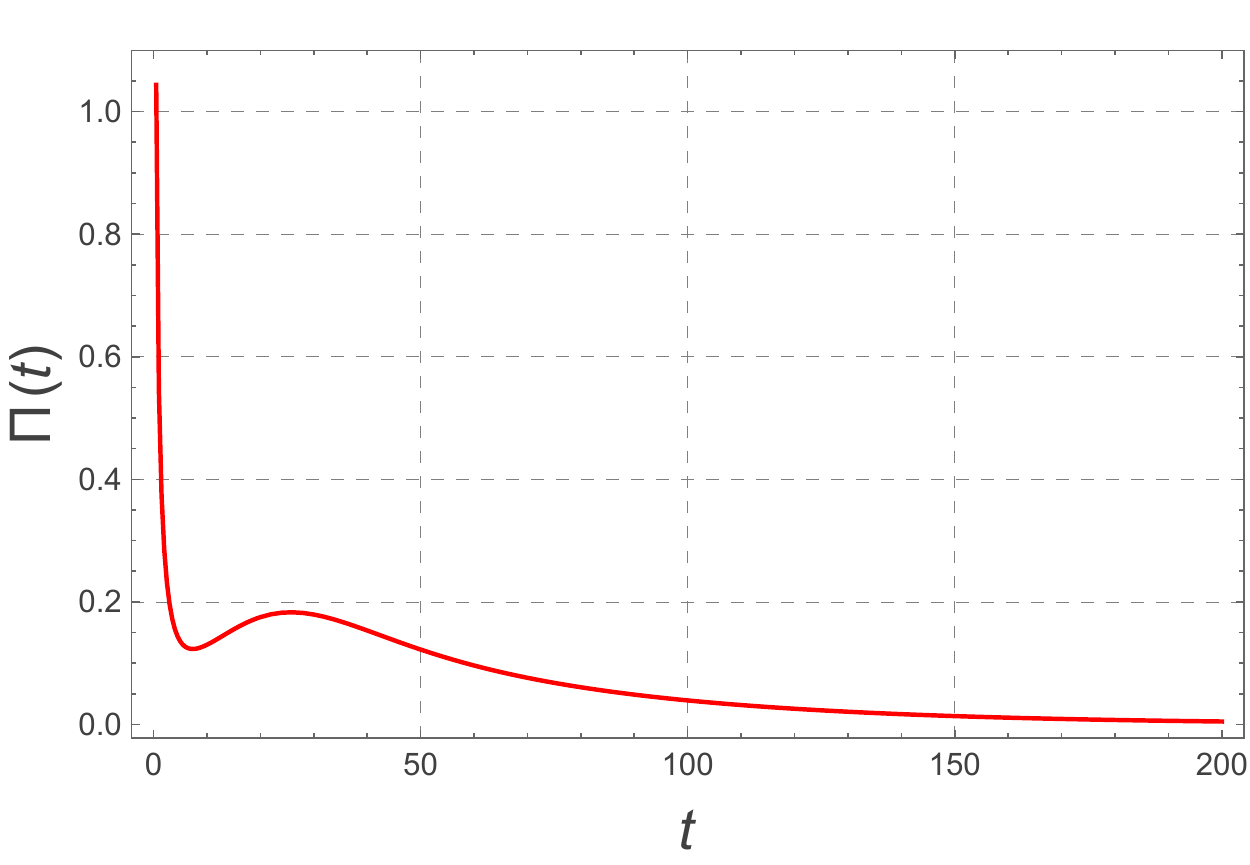}
}
\caption{Entropic evolution initialized at the unstable local maximum.}\label{entropy_unstable}
\end{figure}

Finally, Fig.\ref{entropy_unstable} shows the dynamics starting from the unstable maximum.
The Shannon entropy in Fig.\ref{entropy_unstable}(a) rises sharply as the repulsive potential causes rapid packet broadening. Correspondingly, the entropy production rate in  Fig.\ref{entropy_unstable}(b) starts immediately with a massive spike. This high value quantifies the violent dissipation of free energy as the system is expelled from the unstable equilibrium. It indicates that the system is initially in a state of maximum irreversibility, actively converting potential energy into heat flux. As the probability packet slides down into the stable basins and equilibrium is approached, this dissipation rate monotonically decays to zero, signaling the restoration of detailed balance.

\section{Conclusion}\label{V}

In this paper, we have investigated the stochastic dynamics of phase transitions in RN-AdS black holes using the Fokker-Planck formalism. Moving beyond the static analysis of equilibrium thermodynamics, we employed the generalized free energy landscape to treat the horizon radius as a stochastic order parameter governed by the Fokker-Planck equation. This approach has allowed us to map the precise temporal pathways of phase transitions, bridging the gap between static phase diagrams and dynamic non-equilibrium evolution.

Our investigation delineated distinct dynamical regimes driven by fundamental mechanisms of statistical mechanics. First, for systems initialized in the metastable potential well, the fate of the black hole depends on the diffusion intensity. Under weak thermal noise, the system remains kinetically trapped in the metastable state for exponentially long timescales. However, with sufficient diffusion, phase transition occurs. The system overcomes the potential barrier and evolves from the metastable black hole phase into the stable phase. Conversely, for systems initialized at the unstable local maximum, the dynamics are governed by the inherent mechanical instability. Driven by the repulsive force at the peak, the initial departure does not require thermal activation. However, as the wave packet bifurcates, only a fraction of the probability slides directly into the globally stable state. The remaining portion is temporarily captured by the metastable well and must subsequently undergo a thermally activated barrier crossing to complete the ultimate relaxation.

Furthermore, we characterized the non-equilibrium thermodynamics of the transition using the Shannon entropy and the entropy production rate. It is crucial to emphasize that, in this stochastic framework, the Shannon entropy does not account for the internal microscopic degrees of freedom of the black hole; rather, it strictly quantifies the macroscopic uncertainty of the system's probabilistic evolution. As the probability packet evolves, the Shannon entropy peaks exactly when the system crosses the potential barrier, marking the moment of maximum macroscopic unpredictability. Simultaneously, the entropy production rate exhibits a distinct global peak. In the context of stochastic thermodynamics, this entropy production rate peak rigorously identifies the barriercrossing event as the point of maximum thermodynamic dissipation and irreversibility.

Collectively, these results suggest that black hole phase transitions should be viewed not as instantaneous geometric jumps, but as continuous, dissipative stochastic processes driven by intrinsic thermal fluctuations. These insights provide a robust framework for exploring the dynamic universality classes of holographic systems and may yield further clues regarding the non-equilibrium microstructure of black hole horizons. Future work could extend it to non-linear deviation scenarios or incorporate quantum effects like horizon quantum fluctuations, deepening understanding of the intersection of black hole physics, thermodynamics, and statistical mechanics.

 \section*{Acknowledgment}
We are grateful to Zhen-Ming Xu for fruitful discussions. This work is supported by the National Natural Science Foundation of China (Grant Nos.12405071, 12275216, 12247103).

\end{spacing}

\providecommand{\href}[2]{#2}\begingroup
\footnotesize\itemsep=0pt
\providecommand{\eprint}[2][]{\href{http://arxiv.org/abs/#2}{arXiv:#2}}
\providecommand{\doi}[2]{\href{http://dx.doi.org/#2}{#1}}

\endgroup


\begin{thebibliography}{99}
	
	




\bibitem{Wald:1993nt}
R.~M.~Wald,
\emph{Black hole entropy is the Noether charge},
\href{https://doi.org/10.1103/PhysRevD.48.R3427}{\emph{Phys. Rev. D }{\bfseries 48}, R3427 (1993)},
[\href{https://arxiv.org/abs/gr-qc/9307038}{{arXiv:gr-qc/9307038}}].

\bibitem{Iyer:1994ys}
V.~Iyer and R.~M.~Wald,
\emph{Some properties of Noether charge and a proposal for dynamical black hole entropy},
\href{https://doi.org/10.1103/PhysRevD.50.846}{\emph{Phys. Rev. D }{\bfseries 50}, 846 (1994)},
[\href{https://arxiv.org/abs/gr-qc/9403028}{{arXiv:gr-qc/9403028}}].

\bibitem{Jacobson:1993vj}
T.~Jacobson, G.~Kang and R.~C.~Myers,
\emph{On black hole entropy},
\href{https://doi.org/10.1103/PhysRevD.49.6587}{\emph{Phys. Rev. D }{\bfseries 49}, 6587 (1994)},
[\href{https://arxiv.org/abs/gr-qc/9312023}{{arXiv:gr-qc/9312023}}].

\bibitem{Clunan:2004ch}
T.~Clunan, S.~F.~Ross and D.~J.~Smith,
\emph{On Gauss-Bonnet black hole entropy},
\href{https://doi.org/10.1088/0264-9381/21/14/009}{\emph{Class. Quant. Grav. }{\bfseries 21}, 3447-3458 (2004)},
[\href{https://arxiv.org/abs/gr-qc/0402044}{{arXiv:gr-qc/0402044}}].

\bibitem{Kim:2013cor}
W.~Kim, S.~Kulkarni and S.-H.~Yi,
\emph{Quasilocal conserved charges in a covariant theory of gravity},
\href{https://doi.org/10.1103/PhysRevLett.111.081101}{\emph{Phys. Rev. Lett. }{\bfseries 111}, 081101 (2013)},
[\href{https://arxiv.org/abs/1306.2138}{{arXiv:1306.2138}}].



\bibitem{Kastor:2009wy}
D.~Kastor, S.~Ray and J.~Traschen,
\emph{Enthalpy and the Mechanics of AdS Black Holes},
\href{https://doi.org/10.1088/0264-9381/26/19/195011}{\emph{Class. Quant. Grav. }{\bfseries 26}, 195011 (2009)},
[\href{https://arxiv.org/abs/0904.2765}{{arXiv:0904.2765}}].

\bibitem{Dolan:2010ha}
B.~P.~Dolan,
\emph{The cosmological constant and the black hole equation of state},
\href{https://doi.org/10.1088/0264-9381/28/12/125020}{\emph{Class. Quant. Grav. }{\bfseries 28}, 125020 (2011)},
[\href{https://arxiv.org/abs/1008.5023}{{arXiv:1008.5023}}].

\bibitem{Dolan:2011xt}
B.~P.~Dolan,
\emph{Pressure and volume in the first law of black hole thermodynamics},
\href{https://doi.org/10.1088/0264-9381/28/23/235017}{\emph{Class. Quant. Grav. }{\bfseries 28}, 235017 (2011)},
[\href{https://arxiv.org/abs/1106.6260}{{arXiv:1106.6260}}].

\bibitem{Kubiznak:2012wp}
D.~Kubiznak and R.~B.~Mann,
\emph{P-V criticality of charged AdS black holes},
\href{https://doi.org/10.1007/JHEP07(2012)033}{\emph{JHEP }{\bfseries 07}, 033 (2012)},
[\href{https://arxiv.org/abs/1205.0559}{{arXiv:1205.0559}}].






\bibitem{Cvetic:2010jb}
M.~Cvetic, G.~W.~Gibbons, D.~Kubiznak and C.~N.~Pope,
\emph{Black Hole Enthalpy and an Entropy Inequality for the Thermodynamic Volume},
\href{https://doi.org/10.1103/PhysRevD.84.024037}{\emph{Phys. Rev. D }{\bfseries 84}, 024037 (2011)},
[\href{https://arxiv.org/abs/1012.2888}{{arXiv:1012.2888}}].

\bibitem{Gunasekaran:2012dq}
S.~Gunasekaran, R.~B.~Mann and D.~Kubiznak,
\emph{Extended phase space thermodynamics for charged and rotating black holes and Born-Infeld vacuum polarization},
\href{https://doi.org/10.1007/JHEP11(2012)110}{\emph{JHEP }{\bfseries 11}, 110 (2012)},
[\href{https://arxiv.org/abs/1208.6251}{{arXiv:1208.6251}}].

\bibitem{Altamirano:2013ane}
N.~Altamirano, D.~Kubiznak and R.~B.~Mann,
\emph{Reentrant phase transitions in rotating anti{\textendash}de Sitter black holes},
\href{https://doi.org/10.1103/PhysRevD.88.101502}{\emph{Phys. Rev. D }{\bfseries 88}, 101502 (2013)},
[\href{https://arxiv.org/abs/1306.5756}{{arXiv:1306.5756}}].

\bibitem{Wei:2012ui}
S.-W.~Wei and Y.-X.~Liu,
\emph{Critical phenomena and thermodynamic geometry of charged Gauss-Bonnet AdS black holes},
\href{https://doi.org/10.1103/PhysRevD.87.044014}{\emph{Phys. Rev. D }{\bfseries 87}, 044014 (2013)},
[\href{https://arxiv.org/abs/1209.1707}{{arXiv:1209.1707}}].


\bibitem{Visser:2021eqk}
M.~R.~Visser,
\emph{Holographic thermodynamics requires a chemical potential for color},
\href{https://doi.org/10.1103/PhysRevD.105.106014}{\emph{Phys. Rev. D }{\bfseries 105}, 106014 (2022)},
[\href{https://arxiv.org/abs/2101.04145}{{arXiv:2101.04145}}].

\bibitem{Cong:2019fqn}
W.~Cong and R.~B.~Mann,
\emph{Thermodynamic Instabilities of Generalized Exotic BTZ Black Holes},
\href{https://doi.org/10.1007/JHEP11(2019)004}{\emph{JHEP }{\bfseries 11}, 004 (2019)},
[\href{https://arxiv.org/abs/1908.01254}{{arXiv:1908.01254}}].

\bibitem{Gao:2021xtt}
Z.~Gao and L.~Zhao,
\emph{Restricted phase space thermodynamics for AdS black holes via holography},
\href{https://doi.org/10.1088/1361-6382/ac566c}{\emph{Class. Quant. Grav. }{\bfseries 39}, 075019 (2022)},
  [\href{https://arxiv.org/abs/2112.11236}{{\ttfamily arXiv:2112.11236}}].


\bibitem{Wang:2022err}
T.~Wang, Z.~Zhang, X.~Kong and L.~Zhao,
\emph{Topological black holes in Einstein-Maxwell and 4D conformal gravities revisited},
\href{https://doi.org/10.1016/j.nuclphysb.2023.116352}{\emph{Nucl. Phys. B }{\bfseries 995}, 116352 (2023)},
[\href{https://arxiv.org/abs/2211.16904}{{arXiv:2211.16904}}].



\bibitem{Wang:2021cmz}
T.~Wang and L.~Zhao,
\emph{Black hole thermodynamics is extensive with variable Newton constant},
\href{https://doi.org/10.1016/j.physletb.2022.136938}{\emph{Phys. Lett. B }{\bfseries 827}, 136935 (2022)},[\href{https://arxiv.org/abs/2112.11236}{{arXiv:2112.11236}}].



\bibitem{Li:2020khm}
R.~Li and J.~Wang, 
\emph{Thermodynamics and kinetics of Hawking-Page phase transition},
  \href{https://doi.org/10.1103/PhysRevD.102.024085}{\emph{Phys. Rev. D}
  {\bfseries 102} 024085 (2020)}.

\bibitem{Li:2020nsy}
R.~Li, K.~Zhang and J.~Wang, 
\emph{Thermal dynamic phase transition of Reissner-Nordstr{\"o}m Anti-de Sitter black holes on free energy landscape},
  \href{https://doi.org/10.1007/JHEP10(2020)090}{\emph{JHEP} {\bfseries 10} 090 (2020)}, [\href{https://arxiv.org/abs/2008.00495}{{arXiv:2008.00495}}].

\bibitem{Risken:1989}
H.~Risken, 
\emph{The Fokker-Planck Equation: Methods of Solution and Applications}, 2nd ed., Springer (1996).

\bibitem{Wei:2021bdr}
S.~W.~Wei, Y.~X.~Liu and Y.~Q.~Wang, 
\emph{Dynamic properties of thermodynamic phase transition for five-dimensional neutral Gauss-Bonnet AdS black hole on free energy landscape},
  \href{https://doi.org/10.1016/j.nuclphysb.2022.115692}{\emph{Nucl. Phys. B }{\bfseries 976} (2022) 115692},
  [\href{https://arxiv.org/abs/2009.05215}{{arXiv:2009.05215}}].

\bibitem{Li:2022khi}
R.~Li and J.~Wang, 
\emph{Energy and entropy compensation, phase transition and kinetics of four dimensional charged Gauss-Bonnet Anti-de Sitter black holes on the underlying free energy landscape},
  \href{https://doi.org/10.1103/PhysRevD.106.106015}{\emph{Nucl. Phys. B }{\bfseries 976} 115714 (2022)},
  [\href{https://arxiv.org/abs/2012.05424}{{arXiv:2012.05424}}].

\bibitem{Li:2024path}
R.~Li, C.~Liu and J.~Wang,
\emph{Phase space path integral approach to the kinetics of black hole phase transition},
  \href{https://doi.org/10.1103/PhysRevD.110.024079}{\emph{Phys. Rev. D}{\bfseries 110} 024079 (2024)},
  [\href{https://arxiv.org/abs/2401.02260}{{arXiv:2401.02260}}].

\bibitem{Yang:2025def}
R.~Li, K.~Zhang, J.~Yang, R.~B.~Mann and J.~Wang,
\emph{Critical slowing down of black hole phase transition and
  kinetic crossover in supercritical regime},
   \href{https://doi.org/10.1103/z8bq-kvq3}{\emph{Phys. Rev. D}{\bfseries 112} 064004 (2025)},
  [\href{https://arxiv.org/abs/2505.24148}{{arXiv:2505.24148}}].


\bibitem{Xu:2025jrk}
Z.~M.~Xu and R.~B.~Mann,
\emph{Thermodynamic Supercriticality and Complex Phase Diagram for the AdS Black Hole},
\href{https://doi.org/10.1103/c39y-zcz6}{\emph{Phys. Rev. Lett. }{\bfseries 136} no.4, 041402 (2026)},
[\href{https://arxiv.org/abs/2504.05708}{{arXiv:2504.05708}}].


\bibitem{Li:2024cavity}
R.~Li and J.~Wang, 
\emph{Thermodynamics and kinetics of state switching for the asymptotically flat black hole in a cavity},
\href{https://doi.org/10.1140/epjc/s10052-024-13535-6}{\emph{Eur. Phys. J. C }{\bfseries 84} 1152 (2024)},
  [\href{https://arxiv.org/abs/2405.09151}{{arXiv:2405.09151}}].
 
\bibitem{ZNSM}
R. Zwanzig, Nonequilibrium Statistical Mechanics, Oxford University Press (2001).



\bibitem{Kampen}
N. G. V. Kampen, Stochastic Processes in Physics and Chemistry, North Holland Press (2007).





\bibitem{Kramers:1940}
H.~A. Kramers, 
\emph{Brownian motion in a field of force and the diffusion model of chemical reactions},
  \href{https://doi.org/10.1016/S0031-8914(40)90098-2}{\emph{Physica}{\bfseries 7} 284--304 (1940)}.
  
  \bibitem{Patrick:2003}
P.~Cheridito, H.~Kawaguchi and M.~Maejima, 
\emph{Fractional Ornstein-Uhlenbeck processes},
  \href{https://projecteuclid.org/journals/electronic-journal-of-probability/volume-8/issue-none/Fractional-Ornstein-Uhlenbeck-processes/10.1214/EJP.v8-125.full}{\emph{Electron. J. Probab. }1-14 (2003)}.

  \bibitem{Kubiznak:2016qmn}
D.~Kubiznak, R.~B.~Mann and M.~Teo,
\emph{Black hole chemistry: thermodynamics with Lambda},
\href{https://doi.org/10.1088/1361-6382/aa5c69}{\emph{Class. Quant. Grav. }{\bfseries 34} 063001 (2017)},
 [\href{https://arxiv.org/abs/1608.06147}{{arXiv:1608.06147}}].





\end{thebibliography}
\end{document}